\documentclass[referee]{raa}
\usepackage{threeparttable}
\usepackage{enumerate}
\usepackage{enumitem}
\usepackage{CJK}
\usepackage{xcolor}
\usepackage{footmisc} 
\usepackage{lineno}
\usepackage{subcaption}

\usepackage[english]{babel}

\usepackage{graphicx, times}
\usepackage{natbib}
\usepackage{amssymb, amsmath}
\bibpunct{(}{)}{;}{a}{}{,}

\usepackage[colorlinks=true, allcolors=blue]{hyperref}

\title{GSC 08227-00723: An AH Pic-type System with an Unusually Large Positive Superhump Excess — Implications for Tidal Effects}

\titlerunning{GSC~08227-00723: An Unusually Large PSH Excess AH Pic Candidate}

\author{{Xin Li \inst{1}\thanks{{Corresponding author: lixin@bjp.org.cn} \href{https://orcid.org/0000-0001-5879-8762}{ORCID: 0000-0001-5879-8762}}} \and {Yong-Kang Sun \inst{2,3}\thanks{ 
\href{https://orcid.org/0000-0002-3935-2666}{ORCID: 0000-0002-3935-2666}}}}

\institute{
  Beijing Planetarium, Beijing Academy of Science and Technology, Beijing 100044, People’s Republic of China \\
  \and
  National Astronomical Observatory of China, Chinese Academy of Sciences, Beijing, 100101, People’s Republic of China \\
  \and
  School of Astronomy and Space Science, University of Chinese Academy of Sciences, Beijing 100049, People’s Republic of China \\
}

\begin{document}
\maketitle

\abstract{Nova-like variables are high-accretion-rate cataclysmic variables (CVs) that, in contrast to dwarf novae, do not undergo outbursts caused by thermal-viscous instability. However, a small group of nova-likes, classified as AH Pic-type stars, show recurrent small-amplitude outbursts, which are unexpected by the classical disk instability model. The physical mechanisms underlying these outbursts are not clear. In this study, we present a comprehensive time-domain analysis of the CV candidate GSC~08227-00723 using photometric data from ASAS-SN and TESS. The long-term light curves reveal a sequence of low-amplitude, recurrent stunted outbursts with recurrence times ranging from 30 to 50 days. Notably, these outbursts are frequently preceded by precursor brightenings, a feature reminiscent of super-outbursts in SU UMa stars driven by tidal instability. Period analysis of high-cadence TESS data identifies a coherent periodic modulation at 0.297~d, likely the orbital period, and a persistent positive superhump signal at 0.352~d. The latter corresponds to an exceptionally large superhump excess of $\epsilon^+ \approx 0.19$, surpassing typical values seen in CVs. Additionally, we detect short-timescale variability resembling quasi-periodic oscillations in the TESS light curves. Based on the outburst properties and photometric behavior, we classify GSC~08227-00723 as a new member of the AH~Pic subclass of nova-like stars. 
We discuss how tidal effects may be involved in the observed behavior, although the exact mechanism is still unclear.

}

\section{Introduction}\label{sec:Intro}

Cataclysmic variables (CVs) are a type of interacting binary star system, which consists of a white dwarf primary and a low-mass Roche-lobe filling late-type secondary. Their variability spans a wide range of timescales and amplitudes, reflecting the complex dynamics of accretion disks. For comprehensive reviews of CVs, see \cite{warner_1995}. 

Dwarf novae (DNe) and nova-like variables (NLs) are common subtypes of non-magnet CVs. A prominent feature of DNe is their semi-regular outbursts with amplitudes of approximately 2-6 mag, whereas NLs do not experience such eruptions and instead maintain a relatively stable brightness over long timescales. It is generally believed that the reason for such different light variations or outburst behaviors lies in the different mass-transfer rates from the late type secondary to the primary star. In DN systems, the mass-transfer rate is below a critical threshold, allowing matter to accumulate in the accretion disk until a thermal-viscous instability is triggered, leading to a sudden enhancement of mass accretion onto the white dwarf. In contrast, NL systems sustain mass-transfer rates above the critical value, keeping the accretion disk in a hot, high-viscosity state and thereby suppressing DN-type outbursts \citep{2001NewAR..45..449L}.

In addition to the orbital modulation, many CVs exhibit photometric modulations with periods slightly longer or shorter than the orbital period, commonly referred to as positive and negative superhumps, respectively. Positive superhumps (PSHs) are generally interpreted as arising from the prograde apsidal precession of an eccentric accretion disk. Negative superhumps (NSHs), on the other hand, are usually attributed to a tilted or warped accretion disk undergoing retrograde nodal precession with respect to the binary orbital plane. The presence of superhumps therefore provides important clues to the structure and dynamical state of accretion disks in CVs \citep{1991PASP..103..735P,1998PASP..110.1132P,2009MNRAS.398.2110W}.

As a subclass of DNe, SU~UMa systems are characterized by the presence of both normal outbursts and longer, brighter superoutbursts. During superoutbursts they exhibit PSHs, generally attributed to the 3:1 tidal resonance that drives the accretion disk into an eccentric and precessing configuration in systems with low mass ratios, typically $q \lesssim 0.25$ \citep{1988MNRAS.232...35W}. 

In addition to DNe outbursts, low-amplitude 'stunted' outbursts have been reported in several DN and NL systems, 
including Z~Cam, IW~And, and AH~Pic \citep{2025ApJS..277...29H}. Both accretion disk instabilities and rapid variations in the mass transfer rate have been proposed as plausible mechanisms responsible for these events, but no consensus has been reached.

Z~Cam and IW~And systems are generally interpreted as DNe operating near the threshold of thermal stability, exhibiting standstills between outbursts \citep{2019PASJ...71...20K}. 
In contrast, AH~Pic systems constitute a subclass of NL, named after the prototype AH~Pictoris \citep{2001MNRAS.325...89C}. In typical systems, long-lasting sequences of stunted outbursts with amplitudes of about 0.5-1.0 mag and recurrence times of roughly 12-30 days have been observed. \cite{2024ApJ...977..153B} identified seven AH~Pic stars with these characteristics, and discussed the relationship of this subtype to Z~Cam and IW~And systems.

Over the past two decades, large-scale time-domain surveys have revolutionized the study of CV variability by providing long-term and high-cadence photometric coverage. Such datasets facilitate more systematic investigations of recurrent outbursts and short-period variability. In this work, we compile and analyze the long-term survey data for GSC~08227-0072, a promising CV candidate selected from the International Variable Star Index (VSX)\footnote{\url{https://vsx.aavso.org/}}, maintained by the American Association of Variable Star Observers (AAVSO). Section~\ref{sec:data} describes the acquisition and processing of the survey data used in this study. The photometric behavior and variability characteristics of the system are analyzed in Section~\ref{sec:analysis}, with particular emphasis on its stunted outburst properties. A discussion of the results is in Section~\ref{sec:Discussion}. The conclusion is presented in Section~\ref{sec:Conclusion}.

\section{Data and Processing} \label{sec:data}

\begin{figure*} 
\centering
\includegraphics[width=0.5\textwidth]{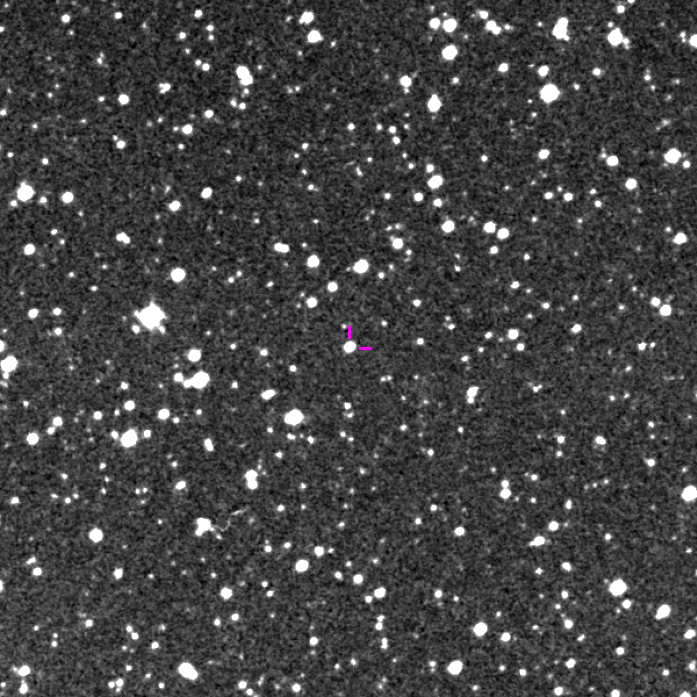}
\caption{GSC~08227-00723 on images of DSS2 survey \citep{DSS2} in red band. The top is north, left is east, and the field of view is $10' \times 10'$ approximately. The target is located in the center the image, as marked by `L' lines. 
\label{fig:general}}
\end{figure*}

GSC~08227-00723 is listed as a CV candidate \citep{2023AJ....165..163C} in the VSX. Its astrometric and photometric parameters from Gaia DR3 \citep{GaiaDR3} and archival imaging from major surveys are presented in Table~\ref{tab:parametersdata} and Figure~\ref{fig:general}, respectively. To date, no prior dedicated time-domain study of this object has been reported in the literature. We selected it because it has extensive ASAS-SN data and simultaneous high-cadence TESS observations, which provide an opportunity to study the relationship between the overall outburst properties and short-timescale variability.

\begin{table}

    \makebox[\textwidth][c]{
    \begin{threeparttable}
	\caption{Parameters of GSC~08227-00723 from Gaia DR3}
	\label{tab:parametersdata}
	\footnotesize
	\begin{tabular}{llr}
		\hline
		 Parameter &   & Value\\
		\hline
    R.A.	&$\alpha$ [h:m:s] 	&11:45:41.76\\
    Dec 	&$\delta$ [d:m:s]	&-51:20:21.12\\
    Apparent magnitude	 &$m_{G}$ [mag]	&13.822\\
    BP-RP	&[mag]	&0.482\\
    Parallax	&mas	&1.1283\\
    Proper motions in R.A.	& mas yr$^{-1}$&-12.244\\
    Proper motions in Dec  & mas yr$^{-1}$&1.517\\
		\hline
	\end{tabular}
    \end{threeparttable}
    }
\end{table}

\subsection{ASAS-SN} \label{subsec:ASASSN}

The All Sky Automated Survey for SuperNovae (ASAS-SN; \citealt{Shappee_2014,Kochanek_2017}) is the first automated survey to monitor the entire visible sky nightly, reaching a depth of $V \sim$ 17 mag. Its global network of 24 telescopes provides coverage of both hemispheres. ASAS-SN is particularly effective at detecting CV outbursts thanks to its high cadence and long temporal baseline. We retrieved the light curve of GSC~08227-00723 from the ASAS-SN Sky Patrol\footnote{\url{https://asas-sn.osu.edu/}}, with early $V$-band and later $g$-band photometry data covering for more than nine years. However, due to its limited spatial resolution with a pixel scale of 8\arcsec\ and PSF FWHM of $\sim$15\arcsec, photometry in crowded fields may suffer from flux contamination, potentially suppressing the true outburst amplitude.

\subsection{TESS} \label{subsec:TESS}

The Transiting Exoplanet Survey Satellite (TESS; \citealt{10.1117/1.JATIS.1.1.014003}) uses four wide-field optical CCD cameras to survey nearly the entire sky, focusing on main-sequence dwarf stars with magnitudes ranging from 4 to 17. Importantly, TESS provides nearly continuous photometric data with a 120-second cadence, spanning approximately 27 days per sector without significant interruptions. This means that each field is observed continuously for a period ranging from about one month to one year, depending on the ecliptic latitude. This capability enables the detection of subtle and rapid photometric variations, making TESS particularly well-suited for identifying periodic signals in CVs, such as orbital periods, and for investigating long-timescale variability in their accretion disks
\citep{2022MNRAS.514.4718B,2023MNRAS.519..352B,2023MNRAS.525.1953B,2024ApJS..273....6B,2024AJ....168..121B,2025ApJS..279...48B,2023AJ....166...56L}.

\begin{figure} 
\centering
\includegraphics[width=1.\textwidth]{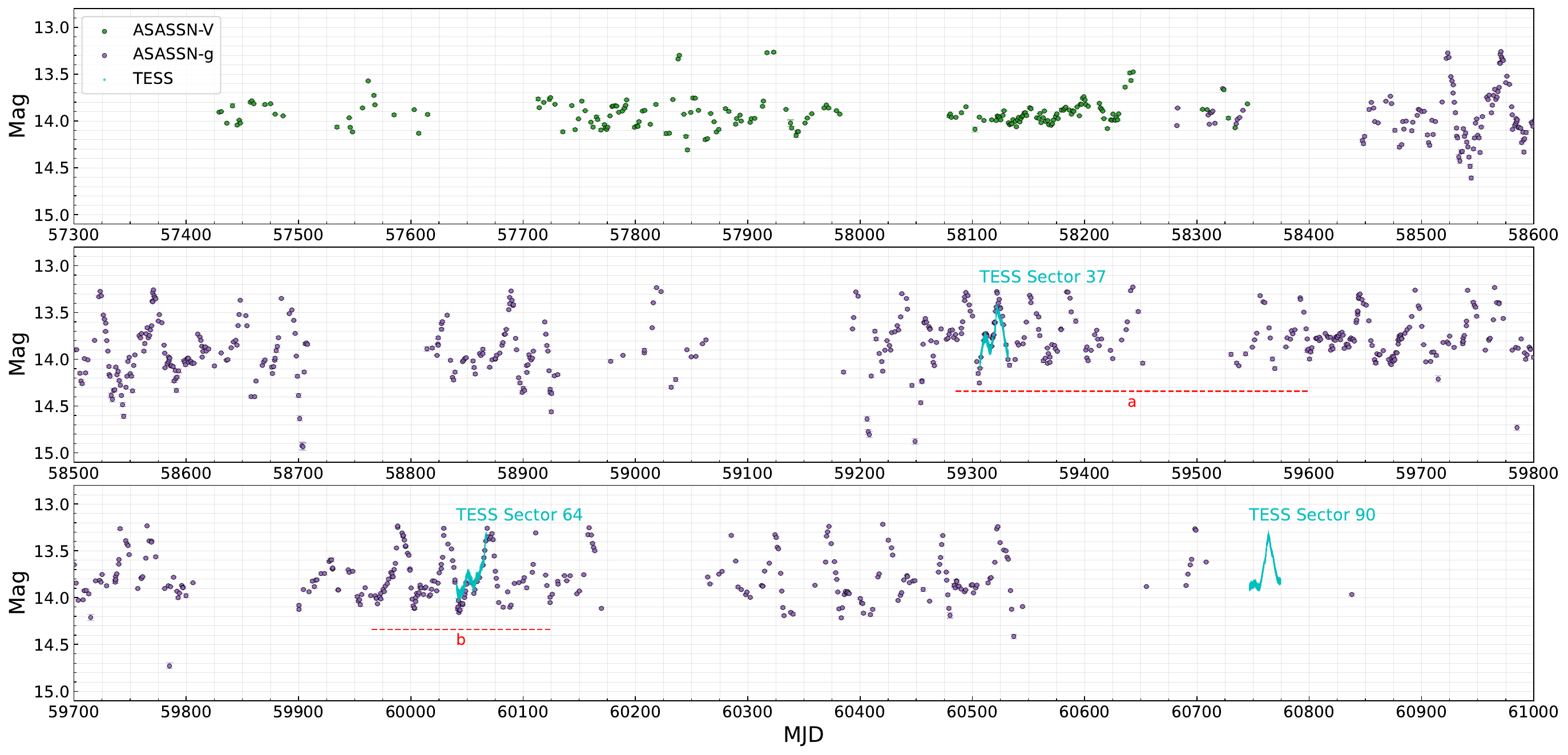}
\caption{ASAS-SN and TESS light curves of GSC~08227-00723. TESS data from sectors 37, 64, and 90 are overlaid on the ASAS-SN data $V$ and $g$-bands. The two segments which show stable outburst recurrence time are labeled as `a' and `b' in red. 
\label{fig:lc}}
\end{figure}

\subsection{TESS light curves of GSC~08227-00723\label{subsec:TESS light curves}}

TESS data from sectors 37, 64, and 90 of GSC~08227-00723 was downloaded through \texttt{Lightkurve} package \citep{2018ascl.soft12013L}. The standard TESS Science Processing Operations Center (SPOC) light curves are optimized for exoplanet detection and employ detrending algorithms to remove instrumental systematics. However, these corrections often inadvertently suppress or eliminate genuine variability on long timescales, such as the overall shape of CV outbursts. To preserve the morphology of the outbursts, we instead constructed light curves directly from the Target Pixel Files (TPFs).

For each of the three TESS sectors, we extracted light curves from the TPFs using the Pixel Level Decorrelation (PLD) method provided by the \texttt{Lightkurve} package. Based on the source positions near the target from Gaia catalog, we defined three apertures required in performing the PLD correction. The first aperture corresponds to the target. The second aperture contains the target along with adjacent background pixels. The third was a background-only aperture that has minimal contribution from any Gaia source. We applied the \texttt{PLDCorrector} to these apertures to remove systematic effects including scattered light and spacecraft-induced variations while retaining the outburst profiles. The resulting 120 s-cadence light curves successfully recover the full shape of the outbursts. We converted the corrected flux to magnitude and manually added a constant offset to align the TESS photometry with the ASAS-SN data. As shown in Figure~\ref{fig:lc}, the TESS outburst profiles are in excellent agreement with the contemporaneous ASAS-SN light curve.

\section{Analysis} \label{sec:analysis}

In this section, we present a detailed analysis of the photometric variability of GSC~08227-00723 based on ASAS-SN and TESS observations. We focus on characterizing its outburst properties, recurrence behavior, and short-timescale periodic signals, with particular attention to features relevant to the AH~Pic subclass of NL.

\subsection{The Outbursts\label{subsec:outburst}}

GSC~08227-00723 exhibits unusual outburst behavior in its ASAS-SN light curve that differs from typical DN eruptions. Prior to MJD 58400, during the period covered by $V$-band observations, the system remained relatively stable with only four visible outbursts while maintaining around $V = 14$ mag with minor fluctuations. Starting around MJD 58500 until the latest data, the system entered a phase of frequent, successive outbursts, reaching a peak brightness of about $g = 13.25$ mag, occasionally followed by dips reaching a minimum brightness of $g \sim 14.9$ mag. Between outbursts, the system returns to a baseline of about $g = 14.1$ mag, yielding an outburst amplitude of $0.85$ mag.

The successive outbursts generally show stable amplitude and maximum flux, with some fluctuations in the recurrence time. To analyze the outburst recurrence time, we calculated the Lomb-Scargle periodogram for different sections of ASAS-SN data. We find that the outburst recurrence time exhibits a drift, varying between approximately 30 and 50 days. Specifically, for the five successive outbursts from MJD 58500 to 58710 the recurrence time was about 41 days, with the third outburst appearing incomplete. Between MJD 59285 and 59600, eight outbursts were recorded with a more stable period of around 29.5 days, which we label segment a, although the fifth outburst seems incomplete. Four consecutive outbursts from MJD 59965 to 60125 show a strong periodicity of 39.2 days, labeled segment b. Finally, from MJD 60300 to the end of observations, the recurrence time lengthened to about 50.4 days. Segments a and b are marked in red in Figure \ref{fig:lc}. Moreover, short-lived quiescent phases during successive outbursts are observed occasionally, such as around MJDs 59625 and 59800.

\begin{figure}[htbp]
    \centering
    \begin{subfigure}{0.48\textwidth}
        \centering
        \includegraphics[width=\linewidth]{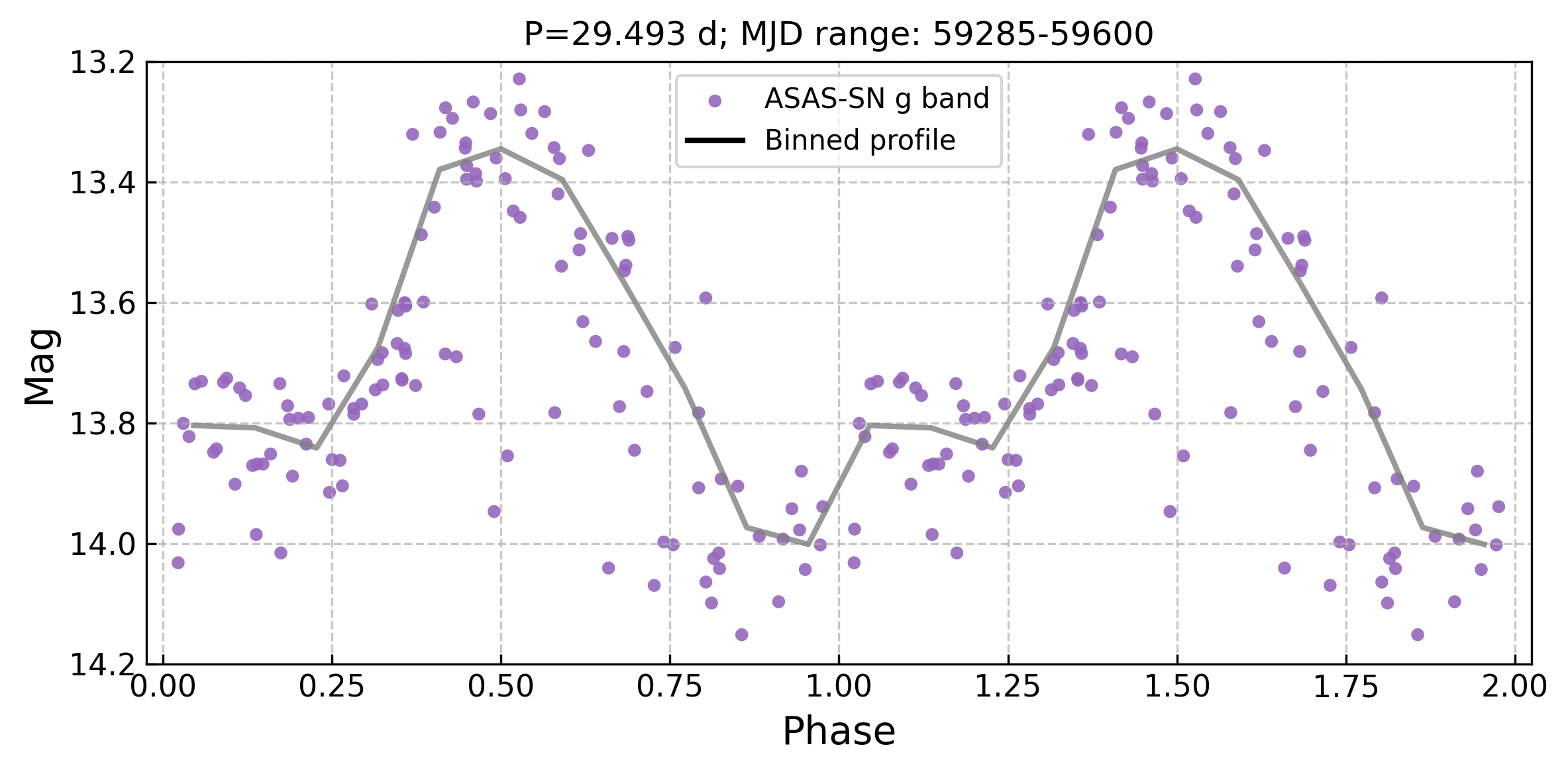}
        \caption{} 
        \label{fig:ob_a} 
    \end{subfigure}
    \hfill
    \begin{subfigure}{0.48\textwidth}
        \centering
        \includegraphics[width=\linewidth]{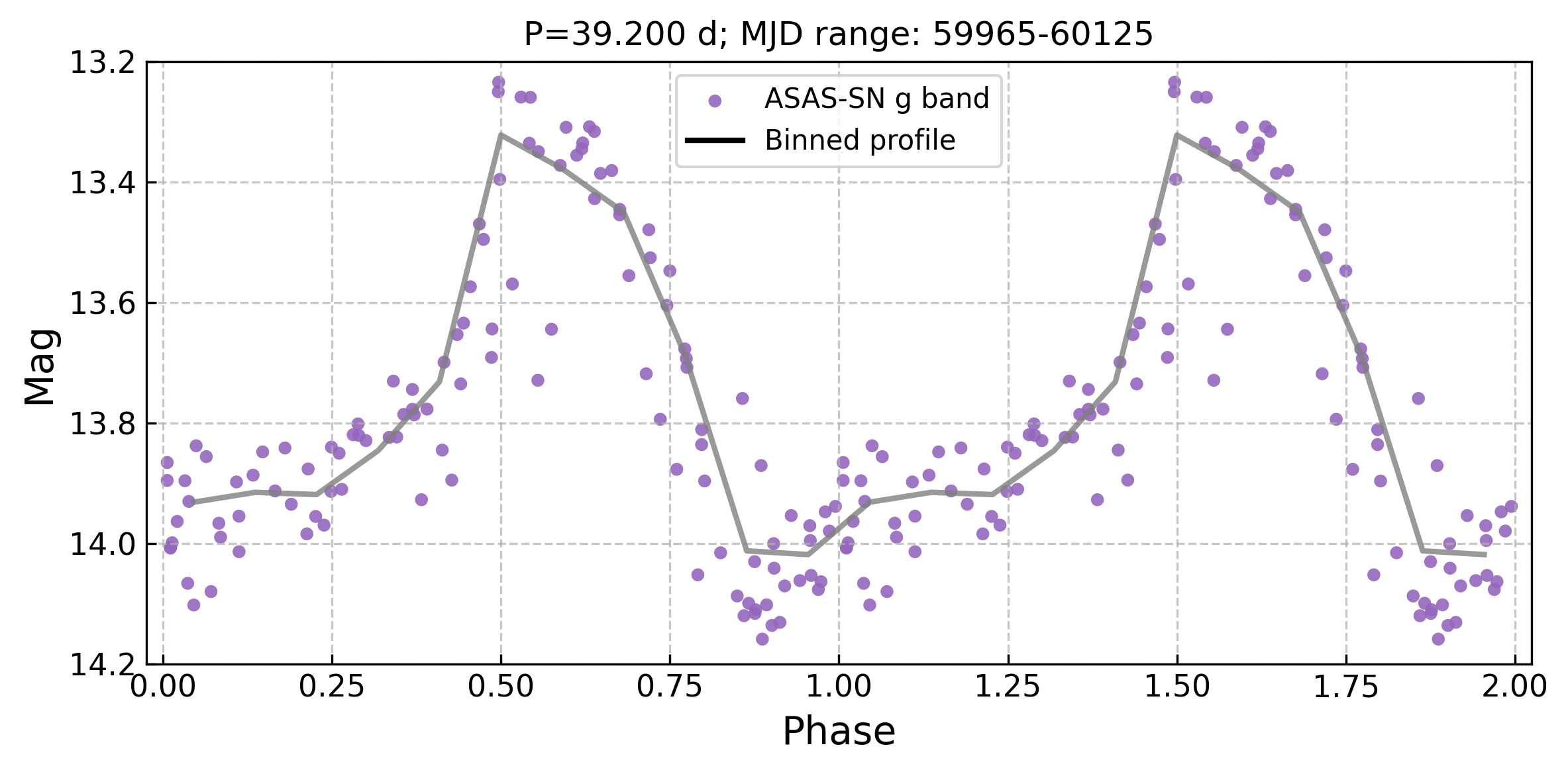}
        \caption{}
        \label{fig:ob_b}
    \end{subfigure}
    \caption{Outburst profiles of GSC~08227-00723. The grey lines are binned data indicating smoothed profiles. Panel (a): Segment a in Figure \ref{fig:lc} between MJD 59285 and 59600, folded with a period of 29.493 d. Panel (b): Segment b in Figure \ref{fig:lc} between MJD 59965 and 60125, folded with a period of 39.200 d.}
    \label{fig:ob} 
\end{figure}

To investigate the outburst profiles despite the insufficient temporal sampling of the ASAS-SN data, we folded the outbursts in segments a and b using their corresponding recurrence periods. The folded results are shown in Figure \ref{fig:ob}, with a smoothed line in each panel indicating the outburst profile. The main feature revealed by both folded profiles is the presence of small precursor brightening events preceding the main outbursts. The peak magnitudes reached in both segments are remarkably consistent. It is notable that near the peak, both profiles exhibit a phase of gradual decline followed by a steeper descent. This two-slope decline is more pronounced in segment b.

\begin{figure}
\centering
\includegraphics[width=.55\textwidth]{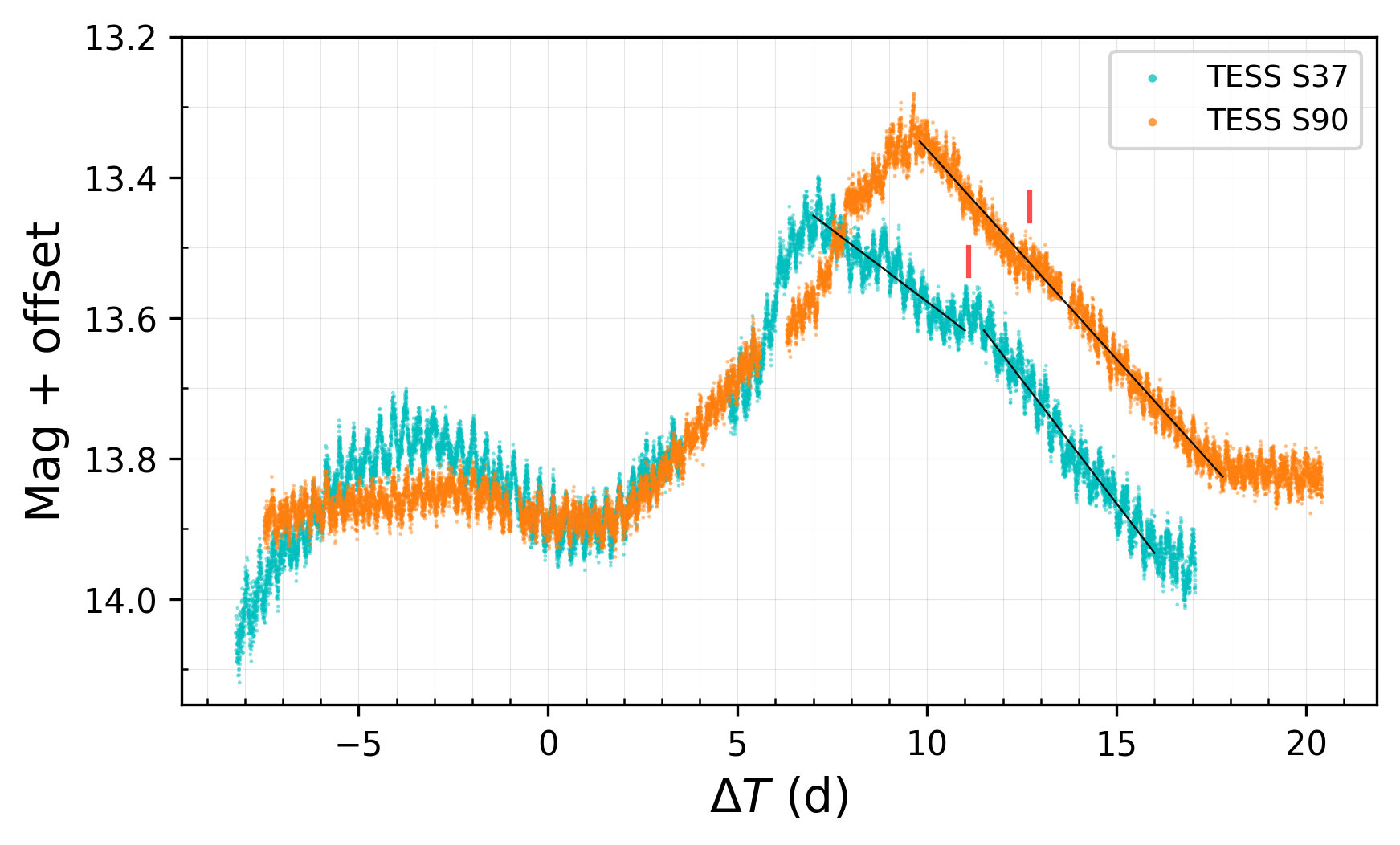}
\caption{Comparison of the two outbursts observed in TESS Sector 37 and 90. The black lines are linear fit to the decline regions. The vertical red lines indicate the rebrightning events after the outburst peaks.
\label{fig:tessob}}
\end{figure}

We compare the two complete outbursts observed by TESS in Sectors 37 and 90, shown in Figure \ref{fig:tessob}, to examine the characteristics of their decline profiles. To establish a common reference point for comparison, we manually align the post-precursor dip, which is a practical choice and does not necessarily imply that these dips represent the same physical phase in both outbursts. We find that the post-peak decline in Sector 37 exhibits a two-slope structure, confirming the profile seen in segment a, while the decline observed in Sector 90 follows a single slope. The linear decline reveals an exponential decline in flux. Intriguingly, both light curves show a subtle re-brightening feature about 3--4 days after the peak.

\subsection{Periodical signals\label{sebsec:period}}

To search for periodic signals and identify the orbital period, we compute the Lomb-Scargle periodogram \citep{Lomb1976, Scargle1982} for the full light curve of each TESS sector. To avoid the effect of the flux variation caused by outbursts, we first flattened the light curves using a Savitzky-Golay filter \citep{Savitzky1964} with a fixed window length to remove long-term outburst trends while preserving short-term variations. The window length was chosen to be sufficiently large to avoid suppressing the amplitude of periodic signals with periods below 0.5~d. (For Sector 64, we discard the data from the last $\sim$2 days, as this interval corresponds to a rapidly rising segment that cannot be adequately flattened.) After identifying the period corresponding to the strongest peak in the periodogram, we divided the original light curve by a sinusoidal model with that period and recalculated the periodogram (also known as pre‑whitening method \citep{Breger1993}). For each sector, we extracted the periods of the three most prominent peaks.

To estimate the uncertainties of the derived periods, we employ a Monte Carlo approach \citep{VanderPlas2018}. We generate an ensemble of synthetic light curves by adding Gaussian noise with a standard deviation equal to the flux uncertainty, and calculate Lomb-Scargle periodogram for each new light curve. The median and standard deviation of the resulting distribution of the strongest period were taken as the final value and its uncertainty. All significant periodic signals detected in TESS Sectors 37, 64, and 90 are summarized in Table\,\ref{tab:gsc_period}, listed in order of decreasing semi-amplitude. We note that the semi-amplitudes listed in Table~\ref{tab:gsc_period} are only meaningful as relative measures within a given TESS sector and should not be directly compared across sectors, as the potential contribution from contaminating sources may vary among different TESS observations.

\begin{figure}
\centering
\includegraphics[width=.55\textwidth]{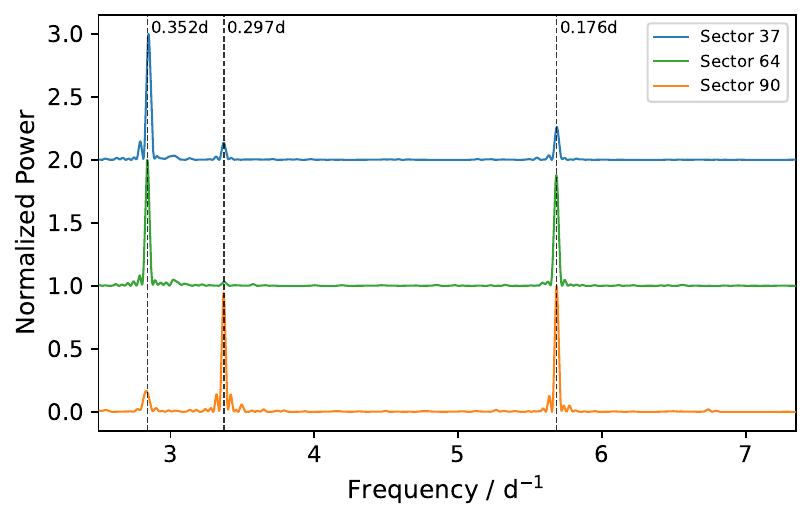} 
\caption{Lomb-Scargle Periodogram calculated from the flattened light curves of the system in TESS Sector 37, 64 and 90. For clarity, the periodograms from different sectors are vertically offset by constant values.
\label{fig:gsc_lsp}}
\end{figure}

The Lomb-Scargle periodogram of the flattened light curves for the three TESS sectors are shown in Figure \ref{fig:gsc_lsp}. The periodograms are normalized by their maximal power. 
It can be seen that the distribution of peak powers differs noticeably among the sectors, reflecting changes in the relative strength of periodic signal components across sectors. Three prominent periods are identified, with periods of \( P_1 = 0.352 \) d, \( P_2 = 0.297 \) d, and \( P_3 = 0.176 \) d.
While \( P_1 \) and \( P_3 \) remain detectable in all three sectors, \( P_2 \) is nearly absent in Sector 64. It is clear that \( P_3\) is the harmonic of \( P_1\) (\( P_3 = P_1/2 \)), while their relative strength exhibit changes. In Sector 37, \( P_1 \) is strong while \( P_3 \) is weak. In Sector 64, \( P_3 \) is nearly as strong as \( P_1 \). In Sector 90, \( P_3 \) is as strong as \( P_1 \). The difference of the peak power distributions reveals the waveform of the short-term variations have significant changes.

\begin{figure}[htbp]
    \centering
    \begin{subfigure}{0.3\textwidth}
        \centering
        \includegraphics[width=\linewidth]{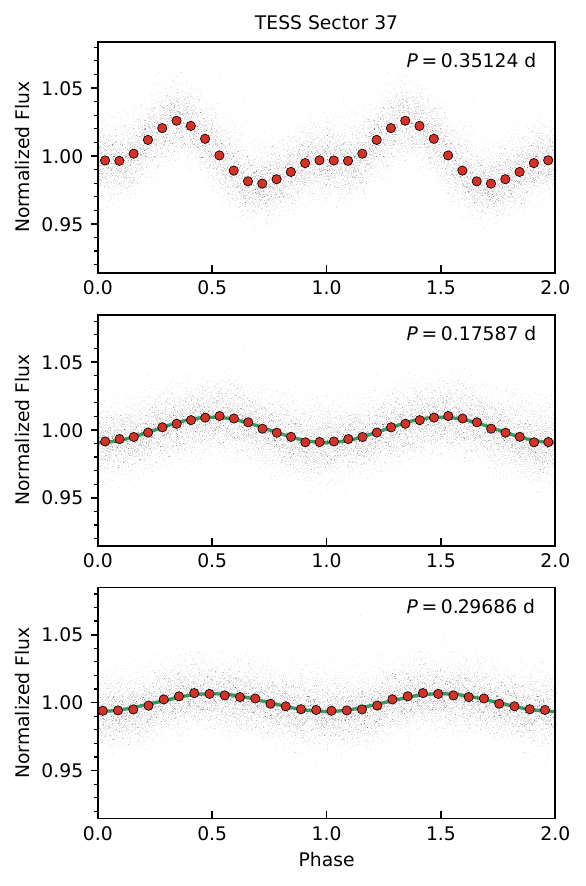}
        \caption{} 
        \label{fig:gsc_tess_S37_fold_a} 
    \end{subfigure}
    \hfill
    \begin{subfigure}{0.3\textwidth}
        \centering
        \includegraphics[width=\linewidth]{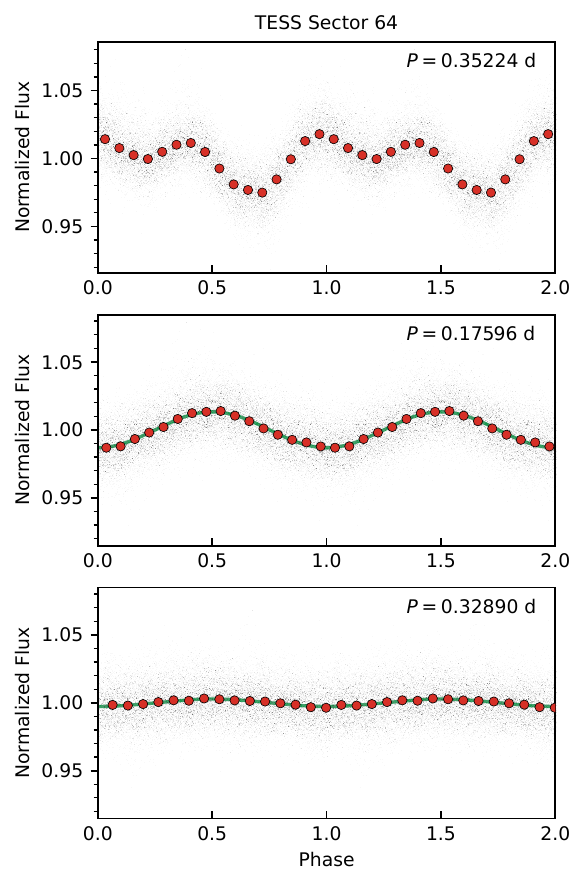}
        \caption{}
        \label{fig:gsc_tess_S64_fold_b}
    \end{subfigure}
    \hfill
    \begin{subfigure}{0.3\textwidth}
        \centering
        \includegraphics[width=\linewidth]{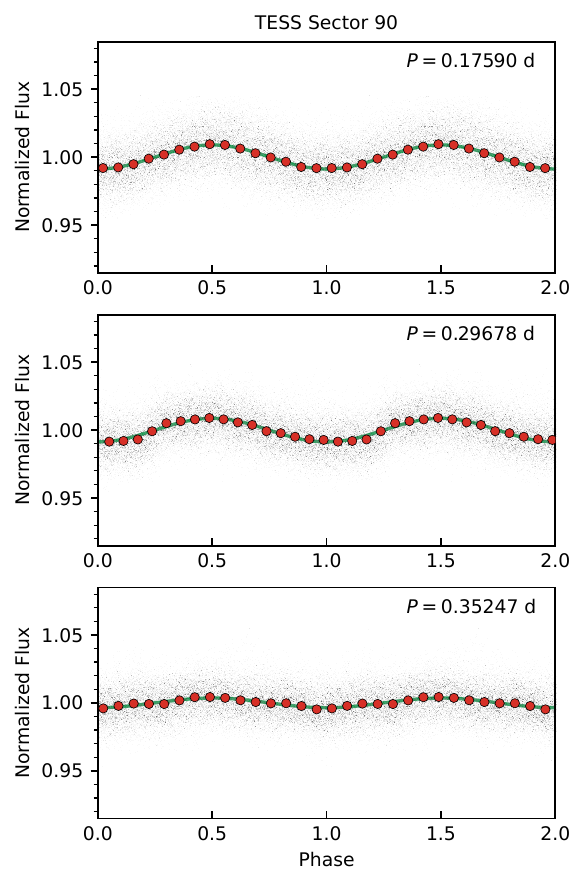}
        \caption{}
        \label{fig:gsc_tess_S90_fold_c}
    \end{subfigure}
    \caption{Phase folded profiles of TESS light curves of the system. Panels (a), (b), and (c) correspond to the results of TESS Sectors 37, 64, and 90, respectively.}
    \label{fig:gsc_tess_fold} 
\end{figure}

We present the phase folded TESS light curves in Figure \ref{fig:gsc_tess_fold}. The three columns (a), (b), and (c) correspond to TESS Sectors 37, 64, and 90, respectively. In each column, the top panel shows the light curve folded on the period of the strongest peak in the Lomb-Scargle periodogram. The middle panel displays the residual light curve after removing the sinusoid model at the first period, folded on the period of the next strongest peak. The bottom panel shows the result of a second pre-whitening step, with the light curve folded on the period of the third strongest peak. 

For Sector 37, the phase-folded light curve using the strongest period of 0.351~d reveals a prominent maximum at one phase and a less pronounced secondary maximum. After removing this 0.351~d signal, the next strongest feature is its harmonic at 0.176~d. Following the removal of both the 0.351~d and 0.176~d signals, a residual periodicity at 0.297~d remains, which is nearly sinusoidal in shape.
For Sector 64, folding the light curve on the strongest dominant 0.352~d period yields a clear minimum and two maxima of nearly equal amplitude. Removal of the 0.351~d signal again uncovers the 0.176~d harmonic as the next strongest component. However, after subtracting both the 0.351~d and 0.176~d signals, the third periodicity is extremely weak and barely discernible.
For Sector 90, the strongest detected period is directly the harmonic at 0.176~d, while the 0.352~d signal appears only as the third strongest, indicating that the modulation of the harmonic by the 0.352~d period is relatively weak in this sector. Meanwhile, the 0.297~d periodicity becomes comparatively more prominent. The dominant period derived from individual sectors lies in the range 0.351-0.352~d and is consistent within uncertainties. For convenience, we adopt $P \approx 0.352$~d hereafter, unless otherwise specified.

Given that the 0.352~d and 0.176~d periods are the strongest in both Sector 37 and Sector 64, we combined the light curves from these two sectors to derive more precise period estimates using the extended temporal baseline. The analysis followed the same methodology applied to individual sectors. Furthermore, we combine all three sectors to obtain the longest available baseline and we find the three period components are consistently significant. The results are presented in the last six rows of the Table~\ref{tab:gsc_period}.

We first consider the possibility that the 0.352~d signal represents the orbital period ($P_{orb}$), while the shorter 0.297~d modulation corresponds to a NSH period ($P_-$).
This interpretation would imply a NSH deficit of
$\epsilon^- = (P_- - P_{orb})/P_{orb} \approx -0.16$, which is significantly larger in magnitude than typically observed values ($ \left| \epsilon^-  \right|\leq 0.05$) in CVs. Such a large deficit would require an unrealistically strong disk tilt and nodal precession rate, making this scenario physically implausible.

We therefore explore an alternative interpretation in which the 0.297~d signal corresponds to the orbital period, while the 0.352~d modulation is identified as a PSH period ($P_+$).
Under this assumption, the inferred PSH excess is $\epsilon^+ = (P_{+} - P_{\rm orb})/{P_{\rm orb}} \approx 0.19$, which is larger than those typically observed in SU~UMa systems, but remains within the range reported for long-period or high–mass-ratio CVs. The 0.176~d signal is naturally explained as the first harmonic of the PSH period, a common feature in superhump-dominated light curves.

Taken together, the extreme NSH deficit implied by the first scenario renders it unlikely. In contrast, the interpretation of the 0.297~d signal as the orbital period and the 0.352~d modulation as a PSH provides a self-consistent explanation of all detected periodicities and yields physically plausible system parameters. We therefore interpret the 0.297~d signal as the orbital period of the system. The 0.352~d modulation is identified as a PSH, while the 0.176~d signal is consistent with the first harmonic of the PSH.

The detected PSH excess places this system among the most extreme cases known in CVs, ranking as the second largest reported to date and only slightly smaller than that of RZ Gru ($\epsilon^+ \approx 0.25$) identified by \cite{2022MNRAS.514.4718B}. If interpreted within the empirical $\epsilon^+$ and mass ratio $q$ relation \citep{2005PASP..117.1204P}, the observed value would correspond to a relatively high mass ratio of $q \sim 0.5$. We note that this estimate is based on an extrapolation beyond the typical range of the relation, and should therefore be treated with caution. We discuss this further in Section \ref{PSH-Period}.

\begin{table*}
    \makebox[\textwidth][c]{
    \begin{threeparttable}
	\caption{Periodic signals from TESS data for GSC~08227-00723}
	\label{tab:gsc_period}
	\footnotesize
	\begin{tabular}{cccc}
		\hline
		 TESS Sector & Period from the original light curve (d) & Period from the MC method (d) & Relative Semi-Amplitude \\
		\hline
        37 & 0.35124112 & 0.351232(17) & 0.0183 \\
		37 & 0.17586925 & 0.1758682(74) & 0.0094 \\
        37 & 0.29685921 & 0.296852(28) & 0.0066 \\
        64 & 0.35224141 & 0.352235(17) & 0.0140 \\
		64 & 0.17596243 & 0.1759604(52) & 0.0133\\
        64 & 0.32890421 & 0.32890(12) & 0.0028\\ 
        90 & 0.17590306 & 0.1759006(74) & 0.0089\\
		90 & 0.29677604 & 0.296774(19)& 0.0088\\
        90 & 0.35246904 & 0.35246(20)& 0.0036 \\
        37+64 & 0.35162468 & 0.35162468(25) & --\\
        37+64 & 0.17591333 & 0.175913329(90) & --\\
        37+64 & 0.29674372 & 0.29674369(94) & --\\
        37+64+90 & 0.35162481 & 0.35162475(30) & --\\
        37+64+90 & 0.17591296 & 0.175912960(55) & --\\
        37+64+90 & 0.29686266 & 0.29674031(18) & --
        \\
		\hline
	\end{tabular}
    \end{threeparttable}
    }
\end{table*}

\subsection{Quasi-Periodic Oscillations\label{QPO}}
\begin{figure}[htbp]
    \centering
    \begin{subfigure}{0.3\textwidth}
        \centering
        \includegraphics[width=\linewidth]{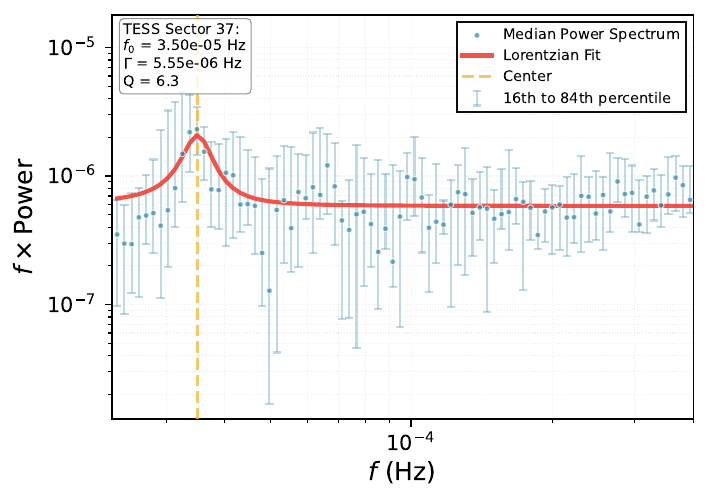}
        \caption{} 
        \label{fig:GSC_TESS_S37_QPO_a} 
    \end{subfigure}
    \hfill
    \begin{subfigure}{0.3\textwidth}
        \centering
        \includegraphics[width=\linewidth]{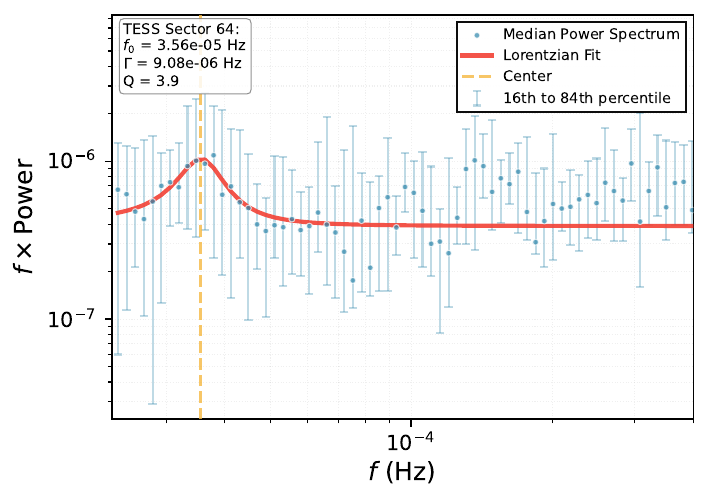}
        \caption{}
        \label{fig:GSC_TESS_S64_QPO_b}
    \end{subfigure}
    \hfill
    \begin{subfigure}{0.3\textwidth}
        \centering
        \includegraphics[width=\linewidth]{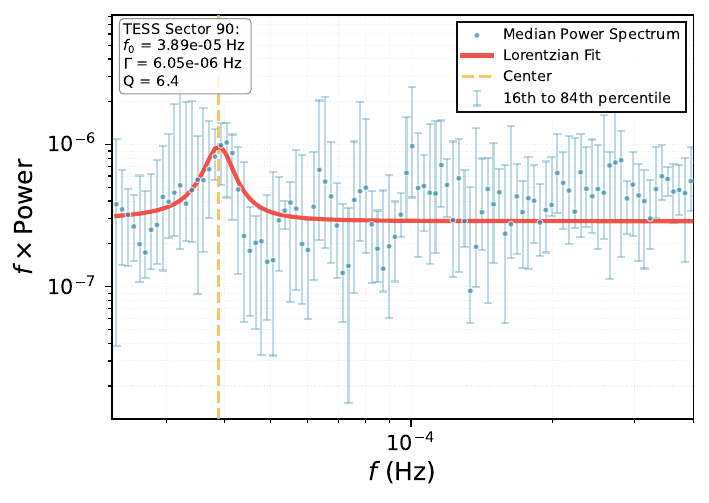}
        \caption{}
        \label{fig:GSC_TESS_S90_QPO_c}
    \end{subfigure}
    \caption{Power spectra calculated after pre-whitening the dominant periodic signals which are analyzed in Sections \ref{subsec:outburst} and \ref{sebsec:period}. The vertical axis shows $f \times Power$, highlighting variability contributions per logarithmic frequency interval. The red lines are models with single Lorentzian plus white noise components fitted to the peak in each power spectrum. The panel--sector correspondence is the same as in Figure~\ref{fig:gsc_tess_fold}. Error bars denote the 16th--84th percentile of the power distribution in each frequency bin.}
    \label{fig:gsc_qpo} 
\end{figure}

After pre-whitening the dominant periodic signals listed in Table~\ref{tab:gsc_period} across all three TESS sectors, we detect a cluster of peaks near 0.3 days, which we suspect to correspond to Quasi-Periodic Oscillations (QPOs; \citealt{Warner_2004}). To investigate this, we divided each TESS sector into multiple contiguous segments. For each segment, we computed the Lomb-Scargle periodogram and transformed the Lomb-Scargle power to f×power units. We then averaged the power within logarithmically spaced frequency bins to rebin each segment's spectrum. The final spectrum was constructed by taking the median f×power value across all segments at each binned frequency.
Uncertainties were determined from the 16th and 84th percentiles of the segment distributions at each frequency bin. The results are shown in Figure~\ref{fig:gsc_qpo}.

To assess the statistical significance of the QPOs, we employ a Monte~Carlo approach based on phase randomization. We first define a frequency window below $2\times10^{-4}$ Hz that encompasses the QPO feature, excluding the band occupied by the Lorentzian peak itself (defined as the signal window). Within this background window, we randomly permute the power values across frequency bins to destroy any coherent signal while preserving the overall power distribution. We repeat this process 5000 times, and we compute the mean excess power as the difference between the average power in the signal window and the average power in the adjacent background window, under the assumption that the underlying noise level is approximately constant over this frequency range. This procedure generates a surrogate power spectrum consistent with the null hypothesis of no periodicity. The fraction of simulations yielding a mean excess power greater than the observed value gives the false-alarm probability, which we convert to an equivalent Gaussian significance using the inverse error function. Finally, we transform the probability to sigma values to show the statistical significance. The resulting significances for TESS Sectors 37, 64, and 90 are 3.1$\sigma$, 2.1$\sigma$, and 1.7$\sigma$, respectively.

The QPO signals are not very significant in the power spectra except for Sector 37. To further confirm the QPO signals, we perform the Weighted Wavelet Z-transform (WWZ) analysis of the pre-whitened light curves. We find drifting signals between frequency 2--6 d$^{-1}$, indicating the existence of QPOs. However, QPOs in CVs are typically observed on timescales of seconds to a few hundred seconds \citep[e.g.,][]{2008AIPC.1054..101W}. A QPO at $\sim 10^{4}$ s would be unusually long and may suggest a distinct physical origin, such as disk precession or thermal-viscous instabilities on large radii.

\section{Discussion} \label{sec:Discussion}

The observed properties of GSC~08227-00723 are not typical of classical long-period NL systems. In this section, we discuss its relation to AH~Pic-type stars and consider the physical implications of its variability.

\subsection{AH~Pictoris Syndrome\label{AH~Pic}}

As shown in Section~\ref{subsec:outburst}, the long-term optical light curve of GSC~08227-00723 exhibits a sequence of low-amplitude, recurrent outbursts that are clearly distinct from DN eruptions. These events are characterized by relatively stable amplitudes, frequent recurrence, and the presence of shallow minima between successive outbursts, a behavior that is particularly evident in the ASAS-SN data. Such phenomenology is consistent with the "stunted outbursts" commonly observed in some NL and long-period CVs.

Within the classification scheme proposed by \cite{2025ApJS..277...29H}, the outburst behavior of GSC~08227-00723 most closely resembles Type 5, which is defined by quasi-continuous sequences of stunted outbursts accompanied by intermittent low states. This pattern matches the defining observational characteristics of the so-called AH~Pic syndrome, as summarized by \cite{2024ApJ...977..153B}, in which systems display long-lasting trains of low-amplitude outbursts over many years without entering a true quiescent state.

In addition to the overall similarity in outburst morphology, GSC~08227-00723 exhibits a feature that has not been previously reported in AH~Pic systems. Specifically, we detect a clear precursor brightening preceding the main stunted outbursts in both the folded ASAS-SN light curves and original TESS time series (Figures \ref{fig:ob} and \ref{fig:tessob}). While precursor outbursts are well known in DNe and are often interpreted within the framework of thermal disk-instability models, their presence in a NL system undergoing quasi-continuous stunted outbursts is unexpected.

The recurrent low-amplitude outbursts, together with the presence of precursor-like structures, suggest that the accretion disk may undergo more complex structural changes than typically assumed for AH~Pic-type systems. This behavior hints that additional physical mechanisms, possibly involving disk asymmetries or tidal effects, may influence the outburst evolution.

\subsection{The \texorpdfstring{$\epsilon^+$-$P_{\rm orb}$ Relation}{epsilon+-Porb Relation} \label{PSH-Period}}

\begin{figure}
\centering
\includegraphics[width=.8\textwidth]{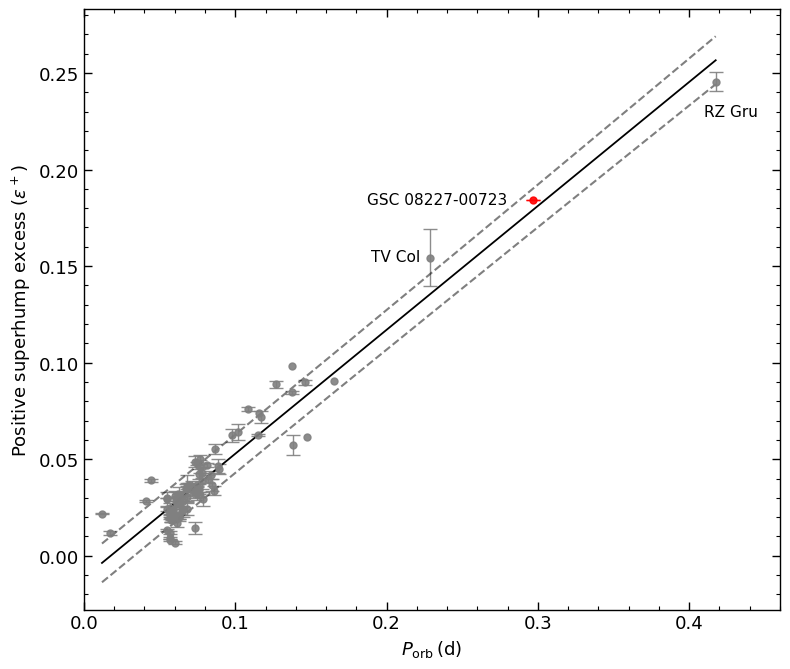} 
\caption{The relation between the PSH period excess ($\epsilon^+$) and the orbital period for superhump systems. Gray circles represent systems compiled by \cite{1998PASP..110.1132P}. The long-period systems TV~Col and RZ~Gru are labeled for reference. GSC~08227-00723 is highlighted in red. For systems with reported uncertainties in $\epsilon^{+}$, error bars are shown. The uncertainty in $\epsilon^{+}$ for GSC~08227-00723 is derived from the propagated uncertainties of the measured periods (see Table~\ref{tab:gsc_period}). The solid line shows the linear fit to the data, and the dashed curves indicate the 1$\sigma$ prediction band of the fit. 
\label{fig:gsc_psh-period}}
\end{figure}

Another notable property of GSC~08227-00723 is its large PSH excess ($\epsilon^+$). Figure \ref{fig:gsc_psh-period} presents the relation between $\epsilon^+$ and orbital period for superhump CV systems, following the compilation of \cite{1998PASP..110.1132P}. The linear fit to the literature sample is shown by the solid line, while the dashed curves indicate the 1$\sigma$ prediction band of the fit, which accounts for both the uncertainty of the regression parameters and the intrinsic scatter of the data.

GSC~08227-00723 is marked in red, which lies on the long-period extension of this relation and falls within the 1$\sigma$ prediction band. Although its superhump excess ($\epsilon^+ \approx 0.19$) is relatively large compared to most NL systems at similar orbital periods, it remains statistically consistent with the empirical trend defined by shorter-period systems. In this sense, it does not represent a clear outlier but rather extends the $\epsilon^+$-$P_{\rm orb}$ relation toward longer orbital periods. The position of GSC~08227-00723 in this diagram suggests that the mechanism responsible for generating PSHs in classical SU~UMa systems may continue to operate, at least phenomenologically, in long-period NL systems. 

We noted that other long-period systems, such as RZ~Gru (its PSHs were tentatively identified by \cite{2022MNRAS.514.4718B} from TESS data) and TV~Col \citep{2003MNRAS.340..679R, 2022Natur.604..447S} marked in Figure \ref{fig:gsc_psh-period}, also exhibit relatively large $\epsilon^+$ at comparable orbital periods. In this context, GSC~08227-00723 appears to belong to this small group of long-period systems that extend the empirical $\epsilon^+$-$P_{\rm orb}$ relation toward higher values of $\epsilon^+$.

As shown in Section~\ref{sebsec:period}, if interpreted within the empirical $\epsilon^+$-$q$ relation of \citet{2005PASP..117.1204P} is applied, the observed $\epsilon^+$ would imply a relatively large mass ratio of $q \sim 0.5$. However, for a CV with an orbital period of $0.297$~d, the semi-empirical donor sequence given by \citet{2011ApJS..194...28K} suggests a donor mass of $M_2 \sim 0.7$--$0.8\,M_\odot$. In this case, such a high $q$ would correspond to an unusually massive white dwarf, approaching or even exceeding the Chandrasekhar limit, which is clearly difficult to reconcile with standard evolutionary expectations.

At such a high $q$, the accretion disk is generally not expected to reach the 3:1 resonance radius under the standard interpretation ($q \lesssim 0.25$), making the presence of a persistent PSH signal less straightforward to explain.

Taken together, these points suggest that the large $\epsilon^+$ observed in GSC~08227-00723 may not translate directly into a reliable estimate of $q$. Instead, it may indicate that the empirical $\epsilon^+$-$q$ relation becomes uncertain in this regime, or that additional physical effects may be involved, as in the case of TV~Col \citep{2003MNRAS.340..679R}. We therefore do not regard the inferred $q$ as a meaningful estimate, and do not attempt to assign it a direct physical interpretation at this stage.

\subsection{Orbital Waveform Evolution and Disk Asymmetry\label{waveform}}

\begin{figure}[htbp]
    \centering
    \begin{subfigure}{0.3\textwidth}
        \centering
        \includegraphics[width=\linewidth]{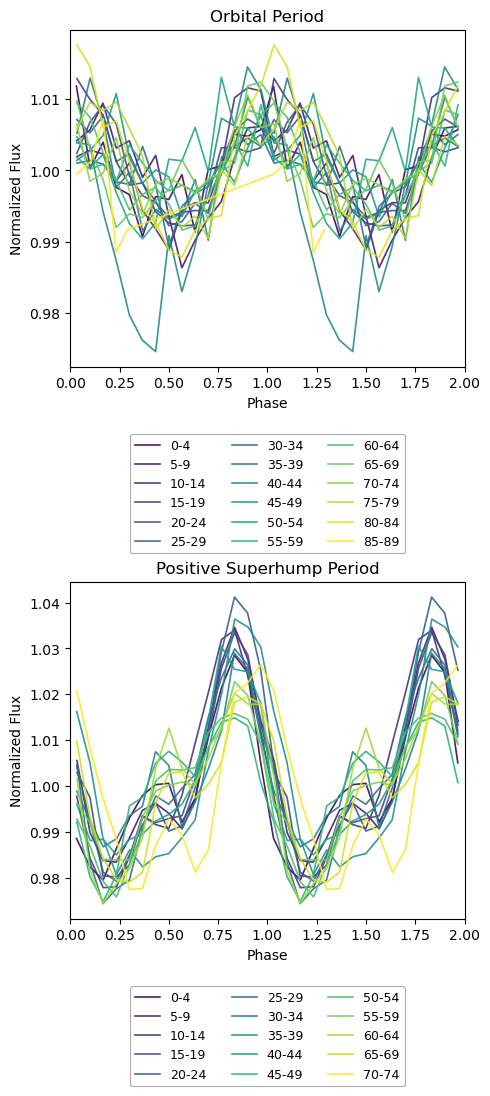}
        \caption{} 
        \label{fig:gsc_profiles_a} 
    \end{subfigure}
    \hfill
    \begin{subfigure}{0.3\textwidth}
        \centering
        \includegraphics[width=\linewidth]{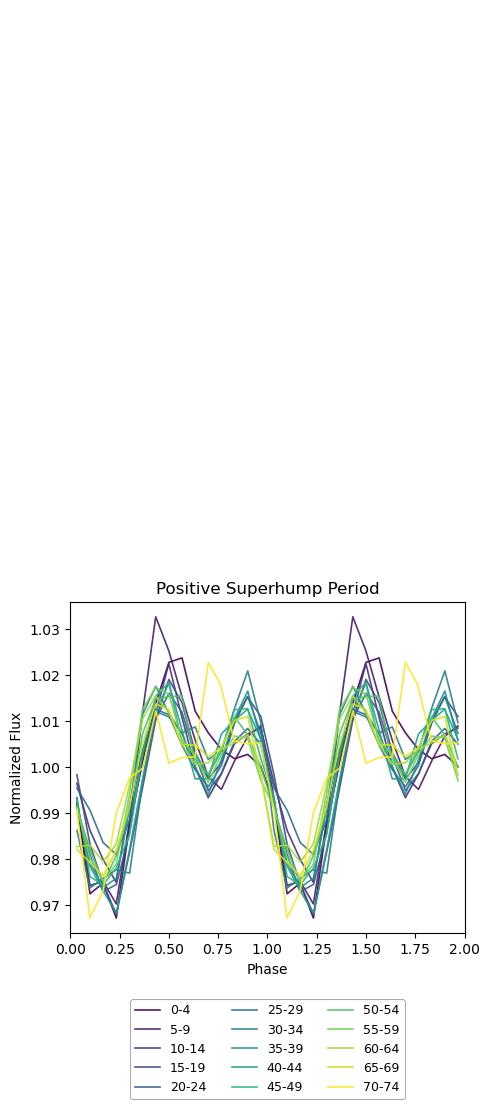}
        \caption{}
        \label{fig:gsc_profiles_b}
    \end{subfigure}
    \hfill
    \begin{subfigure}{0.3\textwidth}
        \centering
        \includegraphics[width=\linewidth]{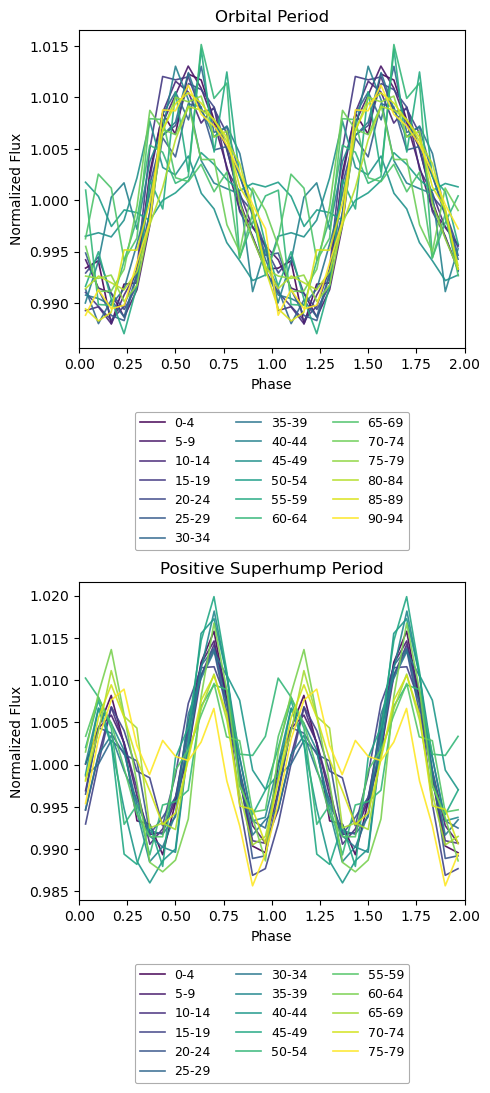}
        \caption{}
        \label{fig:gsc_profiles_c}
    \end{subfigure}
    \caption{Phase profile of the system folded on the orbital (top) and PSH period (bottom) in TESS Sectors 37, 64, and 90 (left to right). For each panel, the data were divided into groups of five consecutive cycles, and each group was phase-folded and binned into 15 phase bins to suppress short-timescale scatter. Different colors represent different five-cycle segments. Two phase cycles are shown for clarity. The flux is normalized by the median flux of each segment.}
    \label{fig:gsc_profiles} 
\end{figure}

The orbital waveform evolution offers additional clues to the disk behavior in this system, complementing the discussion above. To examine the 0.297 d orbital period modulation, we folded the light curves of the available sectors using this period and inspected the stability of the resulting orbital waveforms, shown in Figure \ref{fig:gsc_profiles}. A coherent modulation is clearly visible in Sectors 37 and 90. In contrast, the 0.297 d signal is much weaker in Sector 64 and is not shown in the folded profiles due to its low significance.

In Sectors 37 and 90, the phase of minimum light remains approximately stable, although the depth and overall profile shape vary between consecutive cycle groups. The waveform is therefore not strictly repeatable.  The profiles are relatively smooth and broadly sinusoidal, without sharp ingress and egress features, suggesting that the system is unlikely to be a high-inclination eclipsing binary.

The variable detectability of the 0.297 d signal between sectors, particularly its weakening in Sector 64, indicates that the disk brightness distribution changes over time. If the modulation were produced solely by a rigid stellar eclipse, it would be expected to appear with similar strength in all epochs. Instead, the changing disk structure appears capable of enhancing or suppressing the orbital signal depending on the instantaneous luminosity pattern.

All sectors exhibit a persistent PSH near 0.352 d, shown in Figure \ref{fig:gsc_profiles}. Permanent PSHs in CVs are commonly interpreted as signatures of an eccentric disk undergoing apsidal precession under tidal influence \citep{1991ApJ...381..259L}. Although the present data do not directly constrain the disk radius, the coexistence of a stable superhump and a variable-strength orbital modulation is consistent with a non-axisymmetric, evolving disk.

Taken together, the evolving orbital waveform and the persistent superhump suggest that the 0.297 d signal traces the orbital motion, while the 0.352 d modulation reflects disk precession. 
The orbital modulation is more likely associated with azimuthally asymmetric emission, potentially coupled with mild geometric effects within the accretion disk, rather than a purely stellar eclipse. The waveform evolution provides additional support for a complex accretion disk, and may indicate that GSC~08227-00723 is an unusual system.

\subsection{Outburst properties and possible tidal effects \label{tidal}}

Tidal-instability-induced outbursts are observed in several subclasses of CVs, where tidal effects have been proposed to contribute to the observed variability \citep{1989PASJ...41.1005O, 1996PASP..108...39O}. In such systems, the outburst behavior is often linked to variations in disk stability rather than being solely governed by changes in the mass transfer rate from the secondary star. In this section, we examine the similarities between the outburst phenomenology of GSC~08227-00723 and tidal-instability-related behavior reported in other CVs, including Z~Cam and IW~And systems.

The long-term light curve of GSC~08227-00723 reveals a sequence of recurrent, low-amplitude outbursts interspersed with occasional deep minima, bearing a strong resemblance to the behavior observed in Z~Cam and IW~And systems. The recurrence time of the outbursts is not strictly periodic, drifting between approximately 30 and 50 days. Such characteristics have been reported in systems where the accretion disk is thought to operate close to a critical stability threshold and is susceptible to tidal perturbations.

In addition, several features of the outbursts in GSC~08227-00723 further strengthen the analogy with tidal-instability-driven phenomena. The presence of precursor-like brightening preceding some outbursts, together with a two-stage decline during the decay phase, bears qualitative resemblance to the phenomenology of SU~UMa-type outbursts, despite the system being a long-period NL. These properties may not be fully explained by a simple thermal-viscous instability.

The TESS data provide additional insight through the evolution of the orbital-scale photometric modulation. As discussed in Section \ref{waveform}, the orbital waveforms exhibit significant changes in amplitude and morphology, both between different TESS sectors and within individual sectors on timescales of only a few orbital cycles. Such behavior indicates a dynamically evolving accretion disk rather than a stable, phase-locked superhump associated with a brightened hot spot. This behavior disfavors scenarios dominated by enhanced mass transfer and instead points toward disk-driven variability.

Taken together, the exceptionally large $\epsilon^+$, the evolution of orbital and PSH waveforms, and the outburst phenomenology resembling that of Z~Cam, IW~And, and SU~UMa systems suggest that tidal effects play a role in shaping the observed behavior of GSC~08227-00723. In this context, the AH~Pic-type outbursts observed in this system may represent a possible tidal effects in the accretion disk. Although the observed behavior does not correspond to a classical SU~UMa-type thermal-tidal instability (TTI), and the large $\epsilon^+$ formally suggest a mass ratio above the canonical 3:1 resonance threshold. The persistence of the PSH signal instead suggests that tidal effects may still play a role in the disk evolution, but the exact mechanism is unclear.

\subsection{Comparison on the Gaia Color-Magnitude Diagram \label{CMD}}

\begin{figure}
\centering
\includegraphics[width=.85\textwidth]{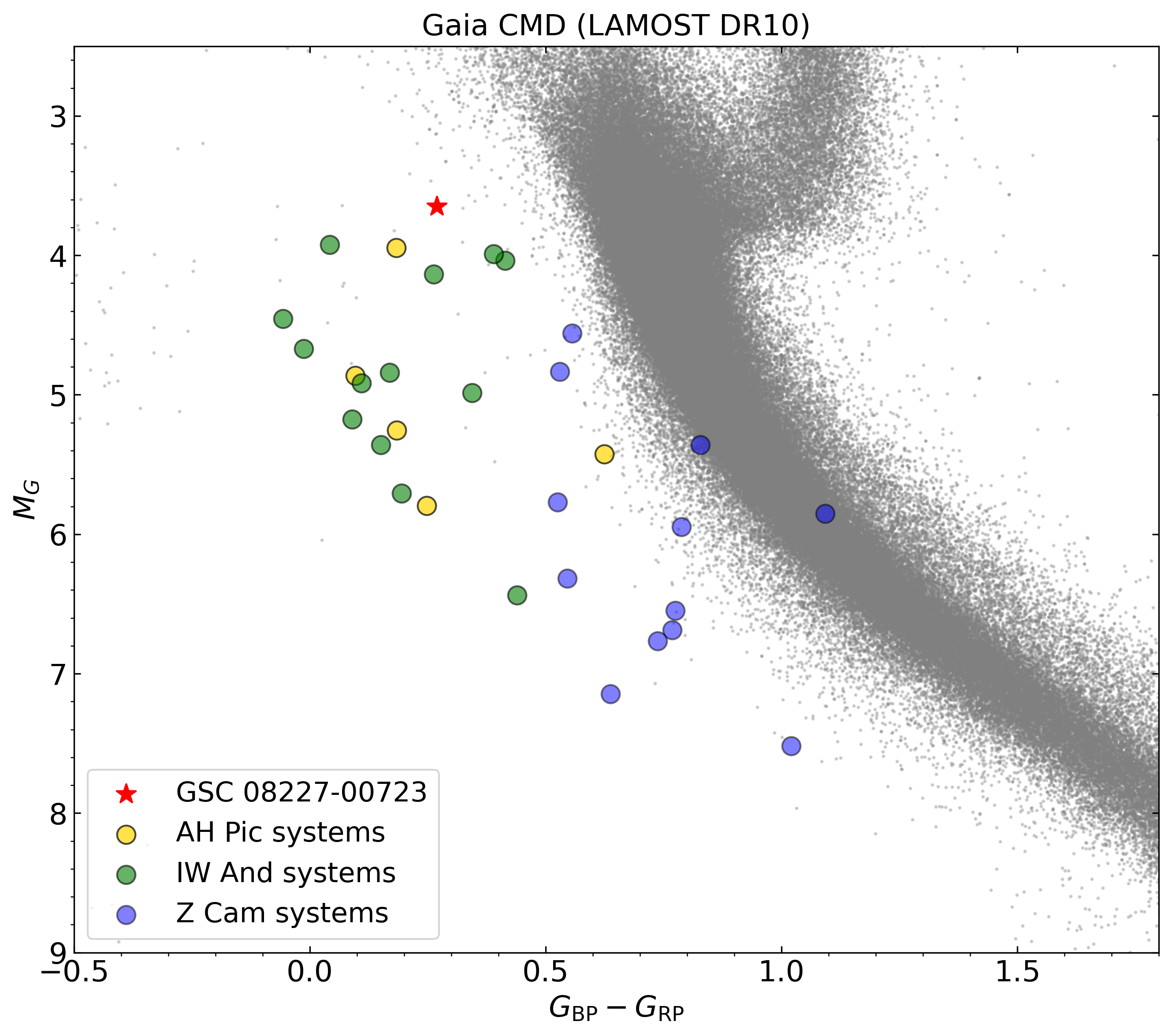} 
\caption{Gaia color--magnitude diagram for the studied systems. The 5 AH~Pic systems, 13 IW~And systems and 12 Z~Cam systems are represented by circles in yellow, green and blue respectively. The main system in this work, GSC~08227-00723, is represented by the red star. The background gray points show field stars randomly selected from the LAMOST DR10 sample. 
\label{fig:gsc_cmd}}
\end{figure}

In Figure \ref{fig:gsc_cmd}, we plotted 5 AH~Pic systems, 13 IW~And systems, and 12 Z~Cam systems on the Gaia colour–magnitude diagram (CMD) as comparison stars. AH~Pic systems which include V1116 Cep, CM Del, V2837 Ori, FY Per, and KIC 9202990 are from \citet{2024ApJ...977..153B}. IW~And systems are IW~And, BC Cas, V513 Cas, BO Cet, V507 Cyg, IM Eri, V523 Lyr, HO Pup, FY Vul, LAMOST J065237.19+243622.1, ASAS J071404+7004.3, Karachurin 12, and KIC 9406652 \citep{Szkody_2013,2019PASJ...71...20K,10.1093/pasj/psab074,2014JAVSO..42..199S,10.1093/pasj/psy138,2022arXiv220305143K,10.1093/pasj/psaa096,2016A&A...589A.106M,2021ApJ...911...51L,2024MNRAS.531..422S,2022MNRAS.510.3605I,2013ApJ...775...64G,2020PASJ...72...94K,2024ApJ...976..107S}. Z~Cam systems are Z~Cam, HS 2325+8205, CRTS J220031.2+033430, AY Psc, BX Pup, EM Cyg, PY Per, RX And, SY Cnc, TZ Per, VW Vul, and WW Cet from \citet{10.1093/mnras/stad2018}. The absolute Gaia magnitudes $M_G$ and $G_{BP} - G_{RP}$ color are calculated from Gaia DR3. The background field stars in the Gaia CMD are randomly selected from the Large Sky Area Multi-Object Fiber Spectroscopic Telescope (LAMOST; \citealt{2012RAA....12.1197C,2012RAA....12..723Z}) DR10 sample. Together with the comparison CVs, they were corrected for extinction using the 3D dust map of \citet{2019ApJ...887...93G} via the \texttt{dustmaps} package. For GSC 08227-00723, the 3D dust map does not provide reliable extinction estimates along its line of sight; thus, we adopted the line-of-sight reddening from the two-dimensional SFD dust map \citep{1998ApJ...500..525S}. The corresponding extinction is $A_G = 0.192$ mag, derived from $E(B-V) = 0.07$ using $A_G = 2.742\,E(B-V)$. The derived extinction is small and does not significantly influence its CMD location.

 In the Gaia CMD, GSC~08227-00723 (marked as a red star) is located at $G_{BP}-G_{RP} = 0.269$ and $M_G = 3.648$. It lies well above the main-sequence locus and within the luminosity range typically occupied by NLs. Compared with other AH Pic systems, GSC 08227-00723 appears slightly more luminous, consistent with emission dominated by a relatively bright accretion disk. Its CMD position overlaps with the bright end of the AH Pic and IW And distributions, while being systematically more luminous than most Z Cam systems in the comparison sample.The combination of a relatively blue color and high luminosity supports a high mass-transfer rate scenario, in agreement with its classification as a long-period NL system exhibiting permanent PSHs. The CMD location therefore reinforces the interpretation that the system resides in a persistently high accretion state, despite showing occasional stunted outbursts reminiscent of DN-like behavior. Given that the optical emission is likely dominated by the accretion disk, the CMD position does not provide a strong quantitative constraint on the contribution from the secondary star.

Taken together, the CMD position, the presence of permanent PSHs, and the AH~Pic-like stunted outbursts are consistent with an AH~Pic classification. In this sense, GSC~08227-00723 may represent a high-luminosity, long-period  example of this subclass with an unusually large PSH excess.

\section{Conclusion} \label{sec:Conclusion}

We present a detailed time-domain study of the AH~Pic candidate GSC~08227-00723 based on ASAS-SN and TESS data. The main findings of this work include:
\begin{itemize}
\item \textbf{Recurrent stunted outbursts.} 
The long-term ASAS-SN light curves reveal a sequence of low-amplitude, recurrent stunted outbursts with recurrence times varying between approximately 30 and 50~days. Several events exhibit precursor-like brightening and a two-stage decline, resembling the phenomenology of SU~UMa--type superoutbursts.

\item \textbf{Orbital and superhump periods.} 
Period analysis of the TESS data identifies a modulation at $P_{\rm orb} \approx 0.297\,\mathrm{d}$ , which varies across different sectors, and a PSH period of $P_{+} \approx 0.352\,\mathrm{d}$.

\item \textbf{Exceptionally large PSH excess.} 
The derived PSH excess $\epsilon^{+} \approx 0.19$ is among the largest known in CVs, second only to that reported for RZ~Gru, placing GSC~08227-00723 at the extreme end of the distribution.

\item \textbf{Evolution of orbital-scale modulation.} 
The TESS light curves reveal significant changes in the amplitude and morphology of the orbital-scale photometric modulation across and within different sectors, indicating a dynamically evolving accretion disk rather than a stable hot-spot--dominated waveform.

\item \textbf{Short-timescale variability.} 
Quasi-periodic oscillations are detected during active phases, further supporting enhanced disk activity.

\item \textbf{CMD position consistent with AH~Pic stars.} 
The Gaia color--magnitude diagram places the system within the luminosity regime of NLs and consistent with AH~Pic systems, supporting its classification as a high mass-transfer system despite exhibiting DN-like outburst behavior.

\end{itemize}

Considering the large PSH excess, stunted outbursts and other light variation characteristics of GSC~08227-00723, we propose that it serves as a significant laboratory for exploring how tidal effects and thermal processes interact.

\section*{Acknowledgments}
This work is supported by the Innovation Project of Beijing Academy of Science and Technology (24CD013). This paper includes data collected by the TESS mission, which are publicly available from the Mikulski Archive for Space Telescopes (MAST). Funding for the TESS mission is provided by NASA's Science Mission Directorate. This research made use of \texttt{Lightkurve}, a Python package for \textit{Kepler} and TESS data analysis \citep{2018ascl.soft12013L}.

\appendix

\section{Some Additional Figures}
\label{sec:Appendix}
 
\setcounter{figure}{0}
\renewcommand{\thefigure}{A\arabic{figure}}

\begin{figure*}
\centering
\includegraphics[width=.8\textwidth]{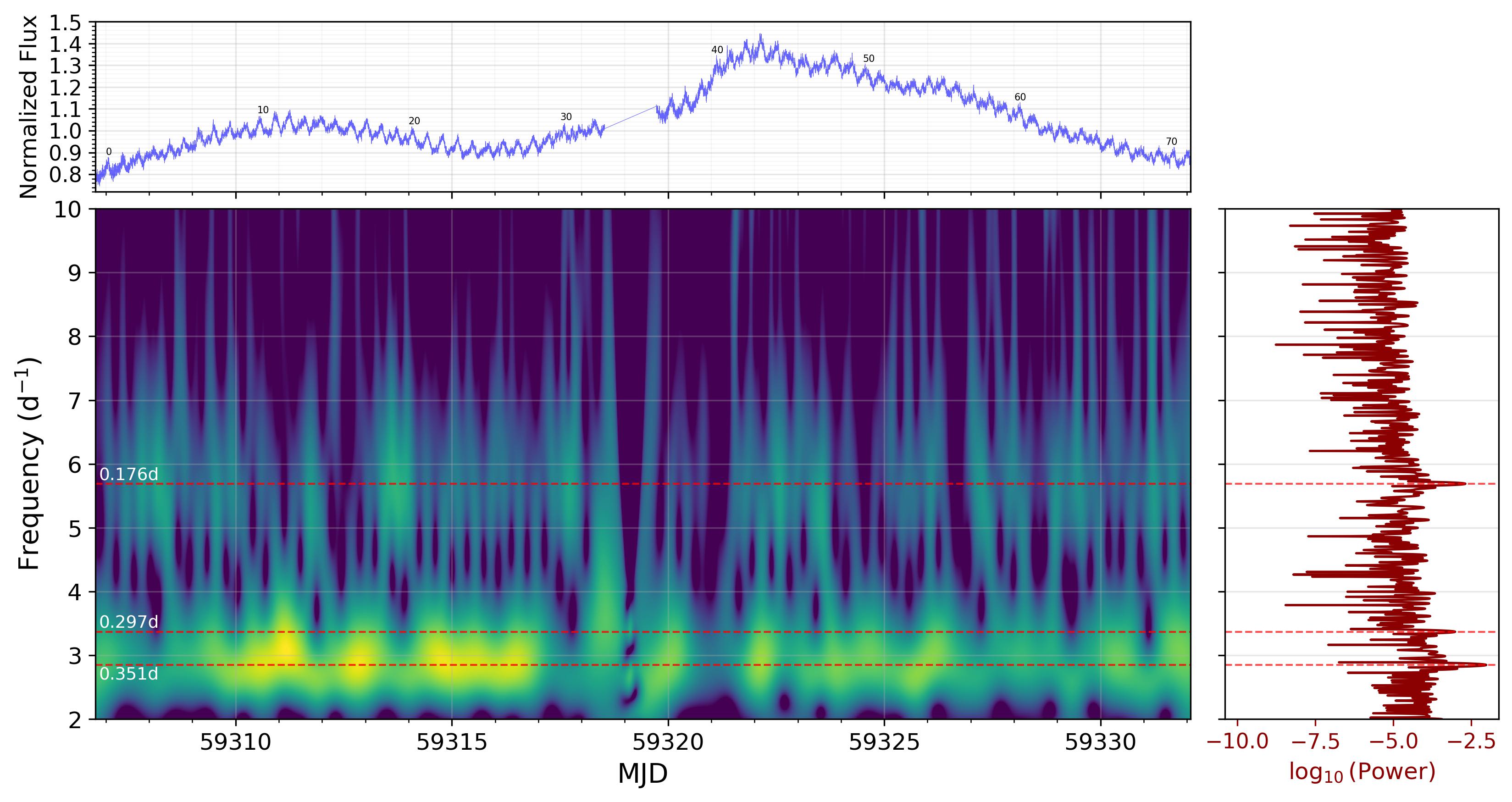} 
\caption{Time–frequency analysis of the TESS Sector 37 light curve of GSC~08227-00723. The top panel shows the normalized flux. The middle panel presents the WWZ power spectrum, and the right panel displays the Lomb–Scargle periodogram of the whole light curve. Dashed horizontal lines indicate three significant periodic signals, which are labeled in the figure.
\label{fig:gsc_S37wwz}}
\end{figure*}

\begin{figure*}
\centering
\includegraphics[width=.8\textwidth]{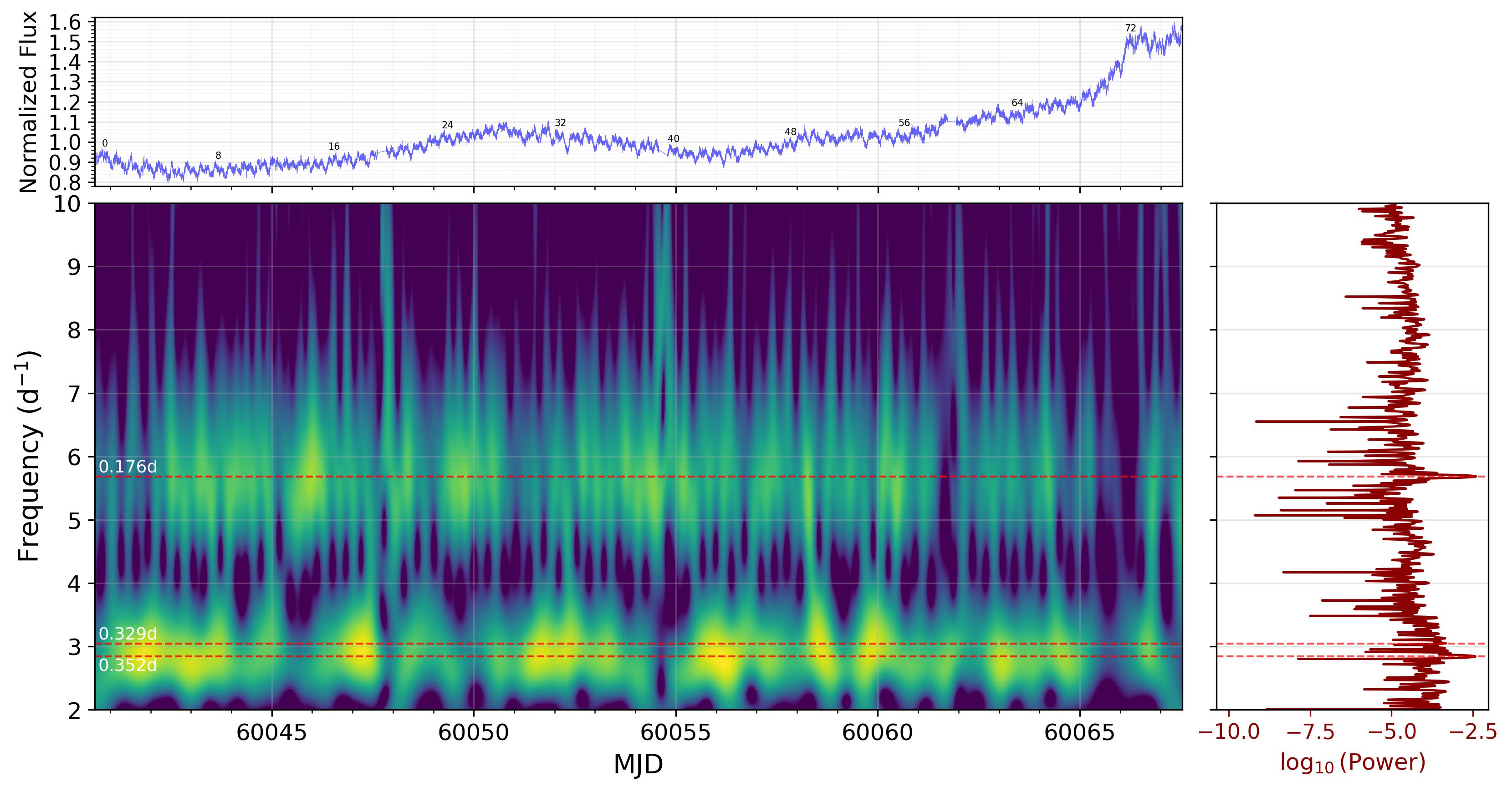} 
\caption{Time–frequency analysis of the TESS Sector 64 light curve of GSC~08227-00723, structured in the same manner as Figure \ref{fig:gsc_S37wwz}. 
\label{fig:gsc_S64wwz}}
\end{figure*}

\begin{figure*}
\centering
\includegraphics[width=.8\textwidth]{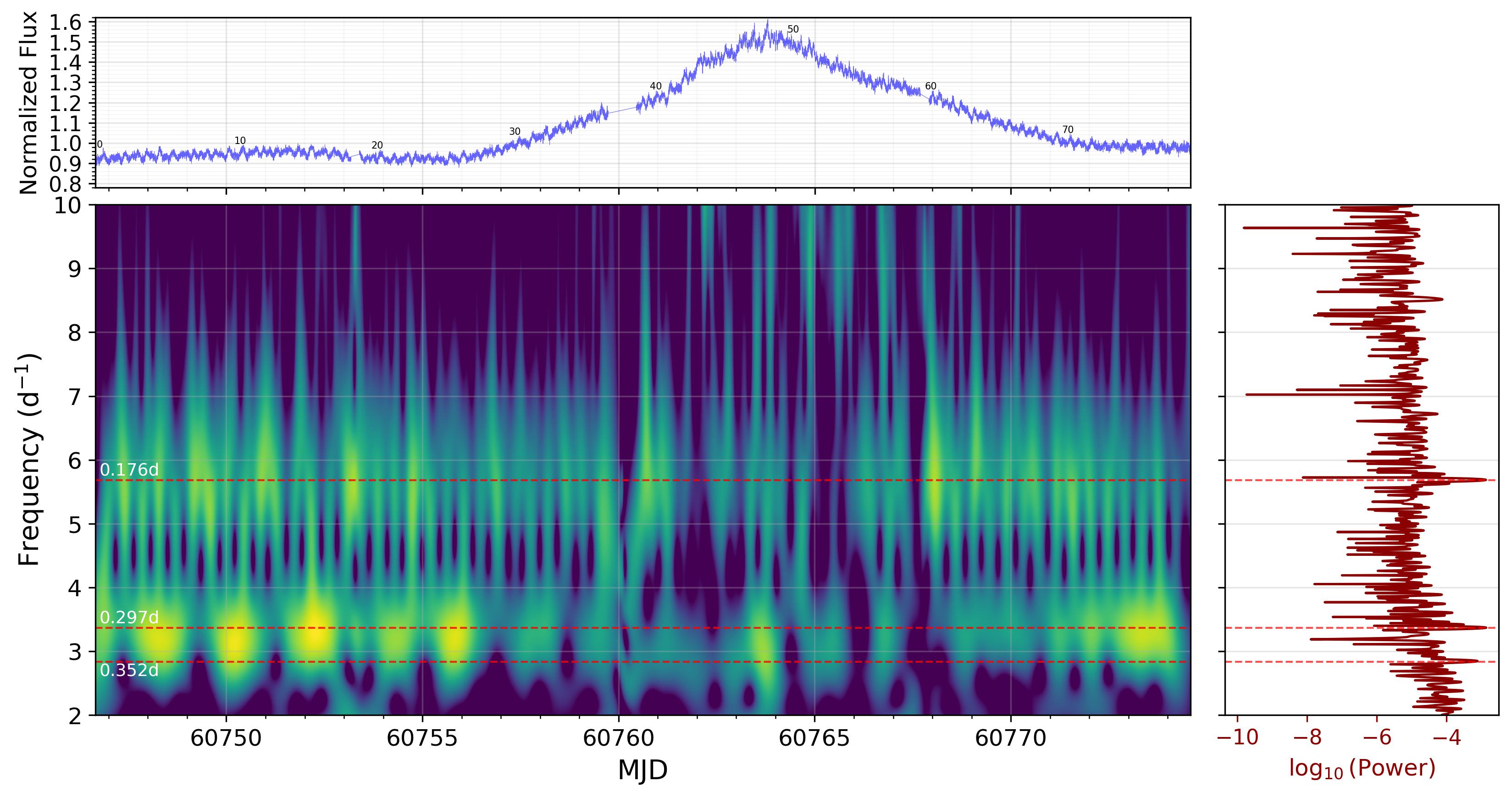} 
\caption{Time–frequency analysis of the TESS Sector 90 light curve of GSC~08227-00723, structured in the same manner as Figure \ref{fig:gsc_S37wwz}.
\label{fig:gsc_S90wwz}}
\end{figure*}

\begin{figure*}
\centering
\includegraphics[width=.8\textwidth]{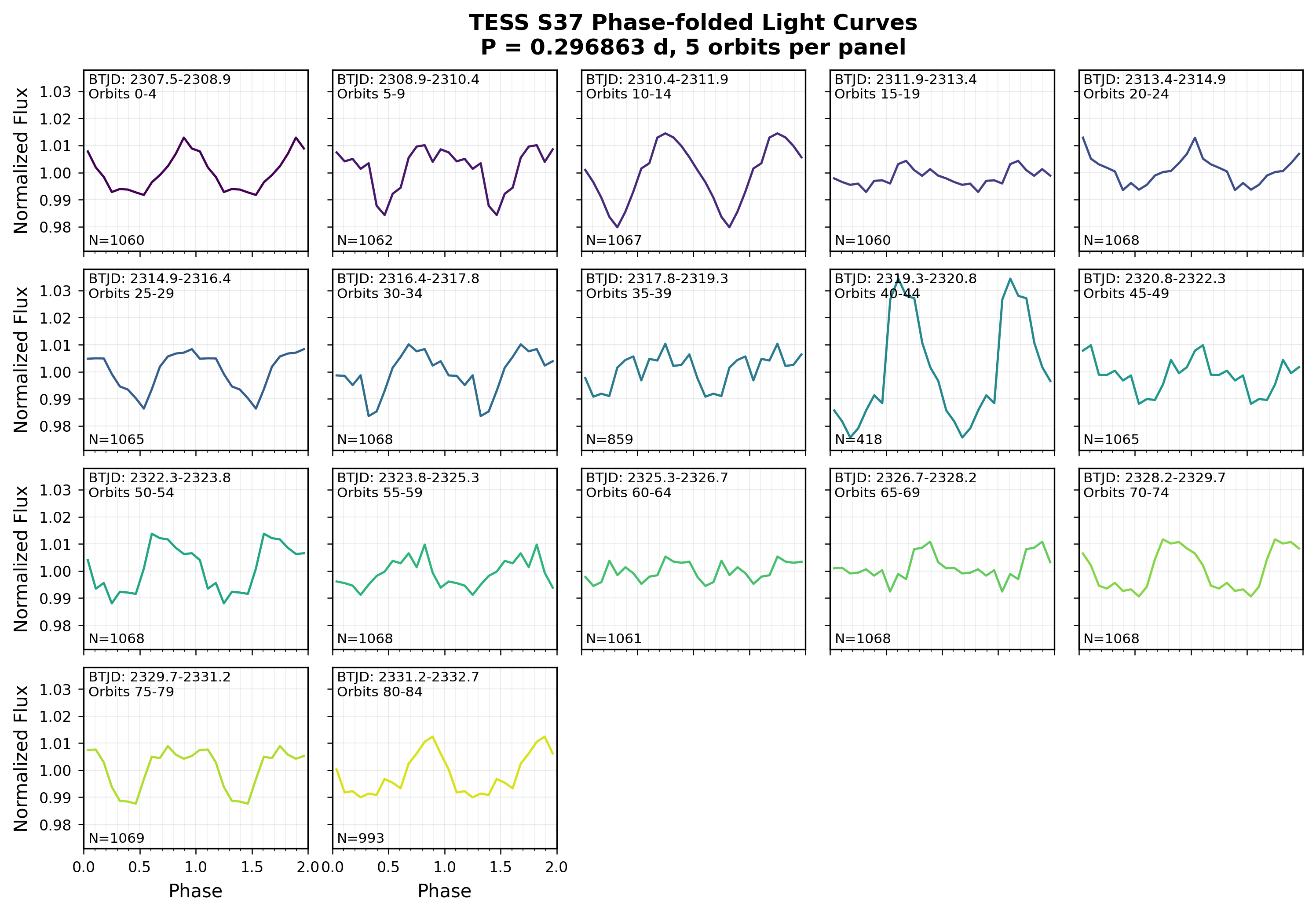}
\caption{Phase-folded light curves of GSC~08227-00723 in TESS Sector 37. The data are folded on the orbital period $P = 0.296863$ d. Each panel corresponds to a consecutive group of five orbital cycles, as indicated in the upper-left corner (BTJD range and orbit numbers). Two orbital cycles (phase 0–2) are shown for clarity. The number of data points included in each panel is indicated by $N$. The sequence of panels illustrates the orbit-to-orbit evolution of the waveform throughout the sector.
\label{fig:gsc_tess37_array1}}
\end{figure*}

\begin{figure*}
\centering
\includegraphics[width=.8\textwidth]{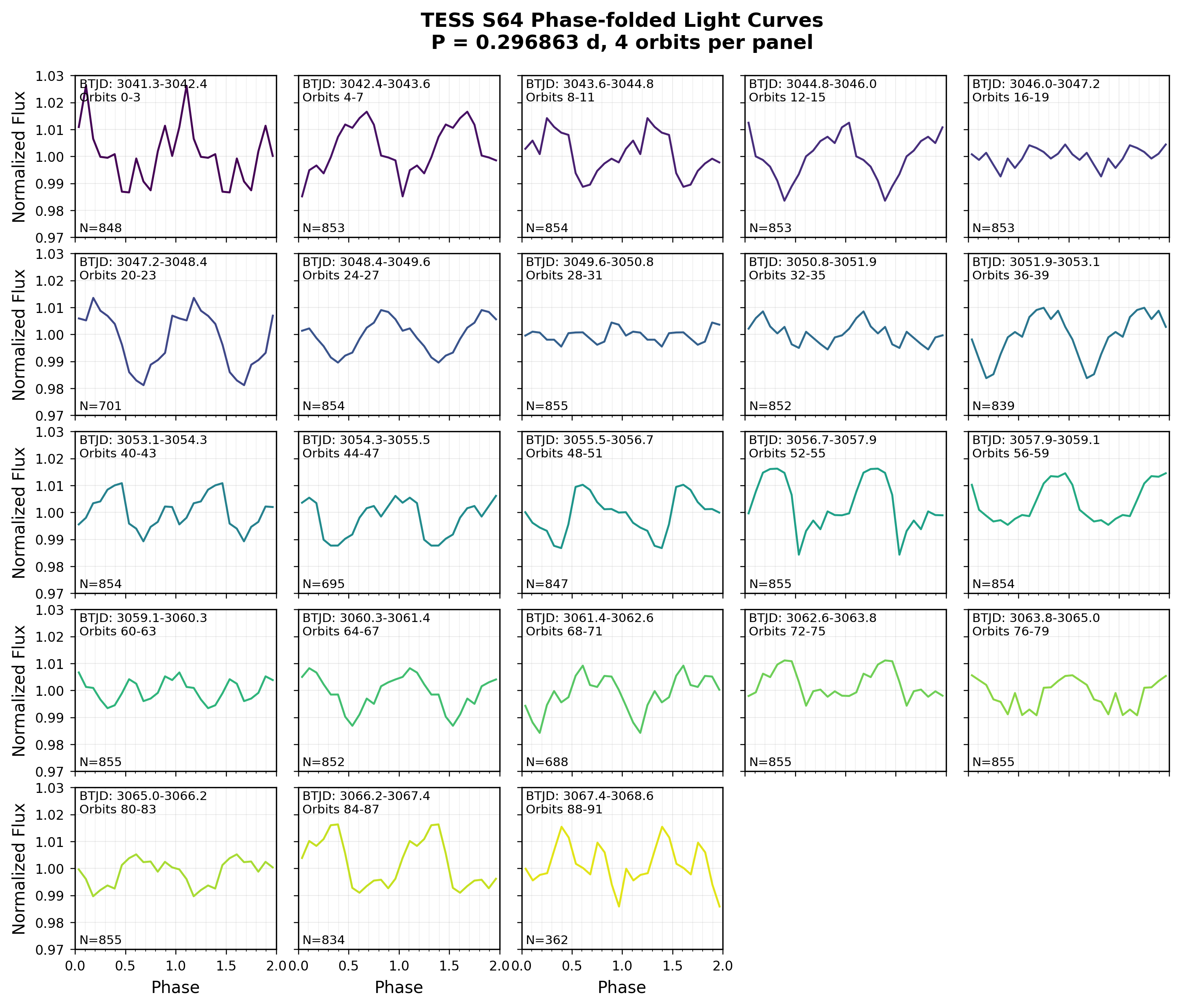}
\caption{Phase-folded light curves of GSC~08227-00723 in TESS Sector 64. Each panel corresponds to a consecutive group of four orbital cycles. Other details are the same as in Figure \ref{fig:gsc_tess37_array1}.
\label{fig:gsc_tess64_array1}}
\end{figure*}

\begin{figure*}
\centering
\includegraphics[width=.8\textwidth]{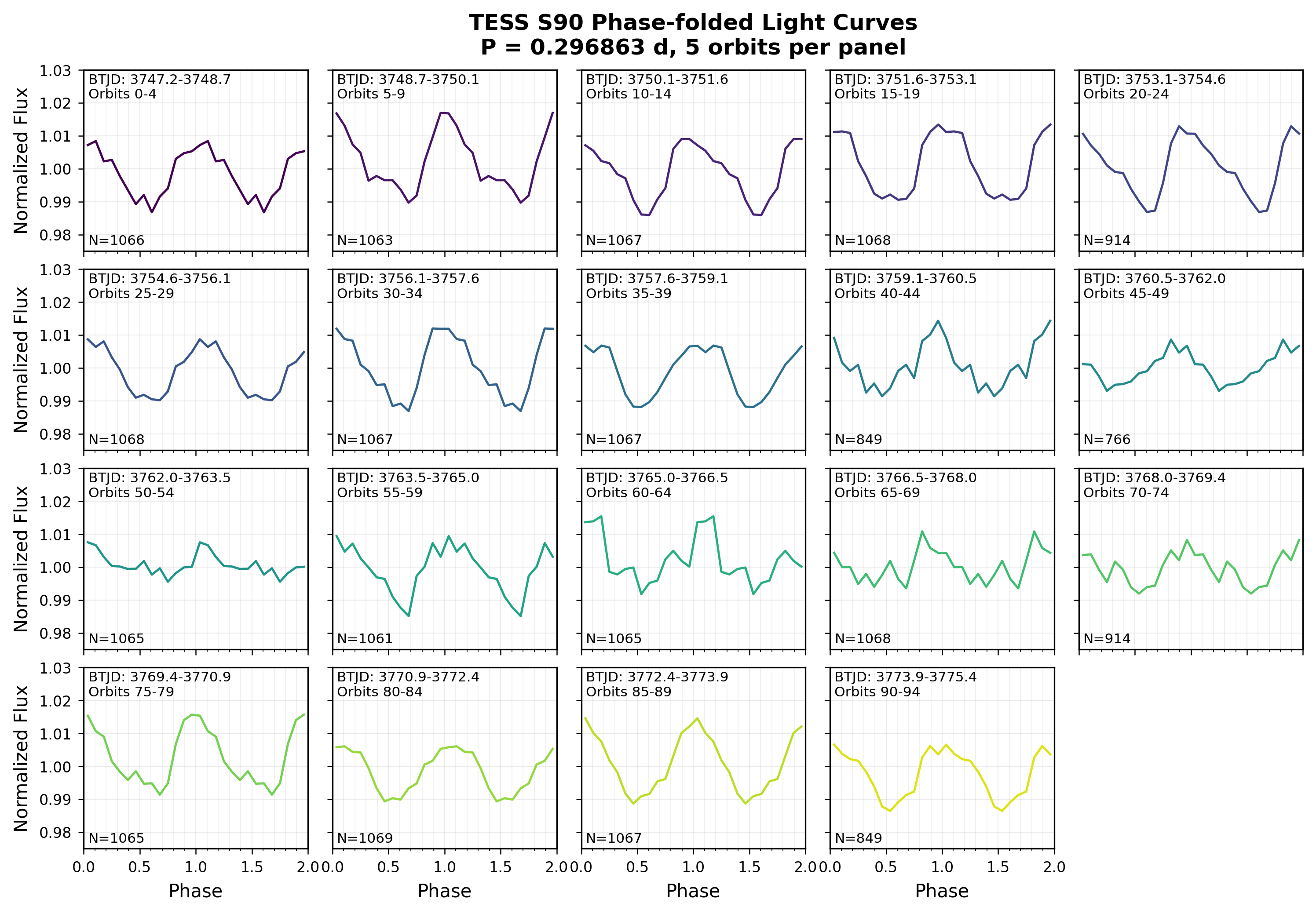}
\caption{Phase-folded light curves of GSC~08227-00723 in TESS Sector 90. Other details are the same as in Figure \ref{fig:gsc_tess37_array1}.
\label{fig:gsc_tess90_array1}}
\end{figure*}

\begin{figure*}
\centering
\includegraphics[width=.8\textwidth]{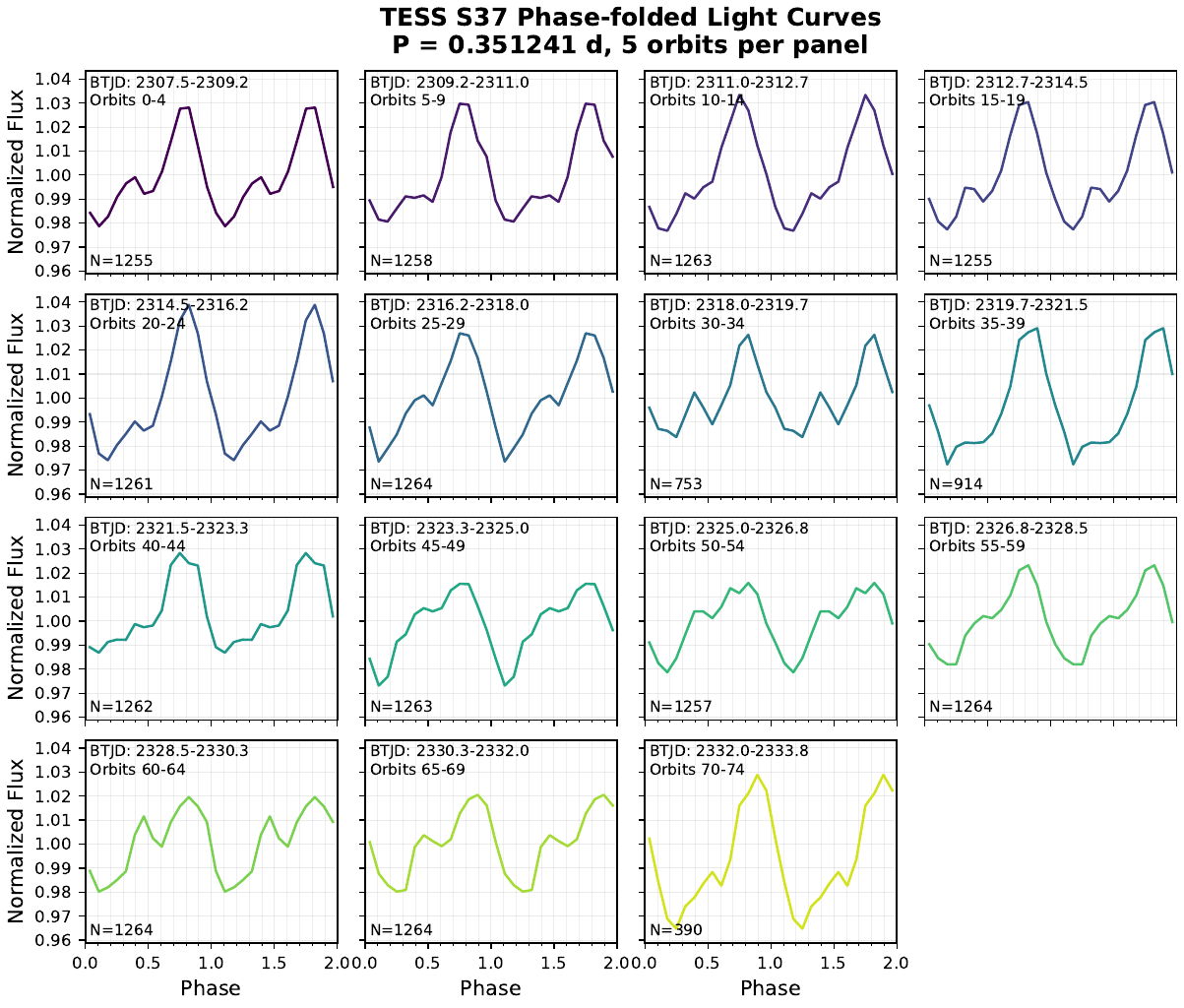}
\caption{Phase-folded light curves of GSC~08227-00723 in TESS Sector 37. The data are folded on the PSH period $P = 0.351241$ d. Other details are the same as in Figure \ref{fig:gsc_tess37_array1}
\label{fig:gsc_tess37_array2}}
\end{figure*}

\begin{figure*}
\centering
\includegraphics[width=.8\textwidth]{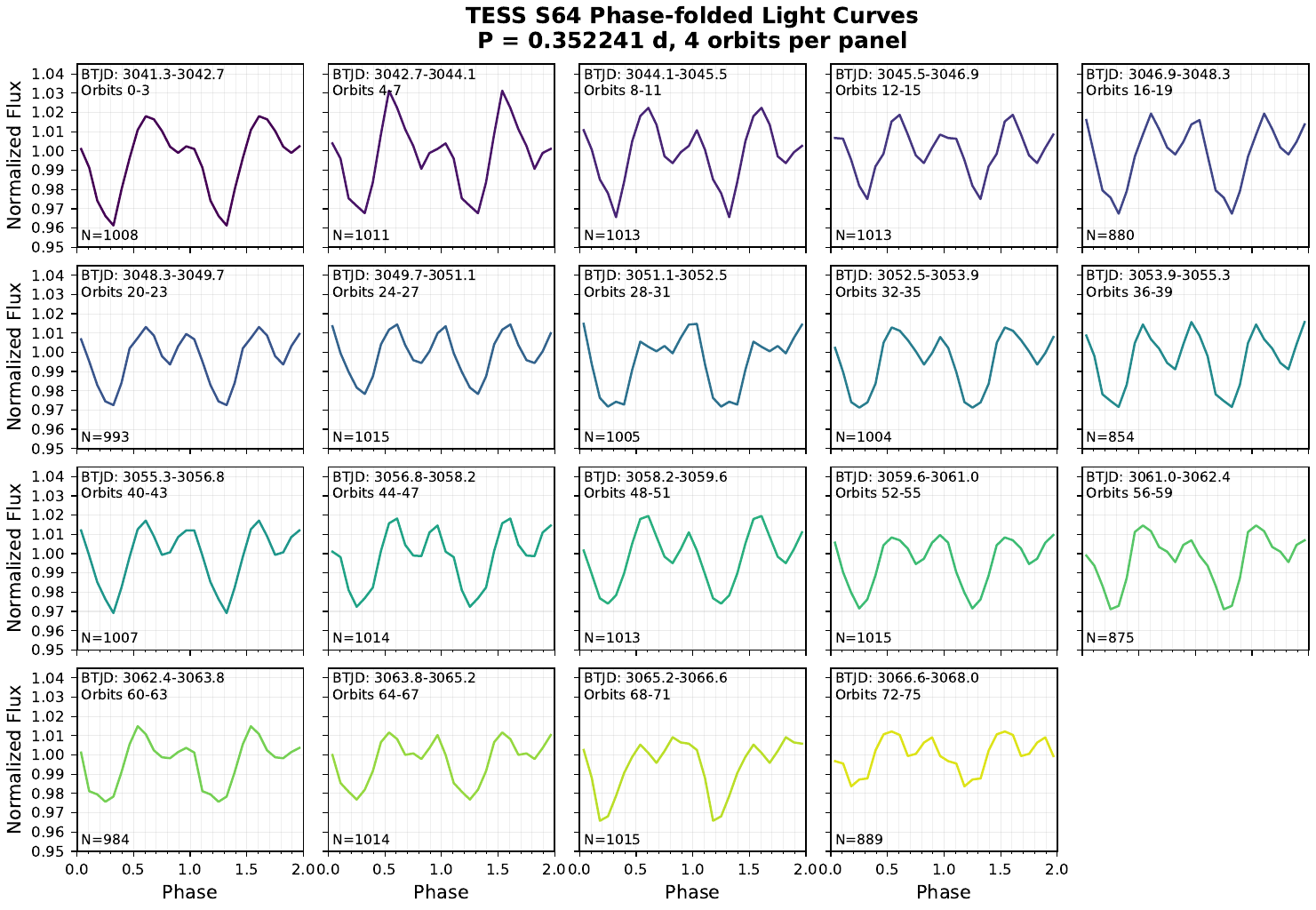}
\caption{Phase-folded light curves of GSC~08227-00723 in TESS Sector 64. Each panel corresponds to a consecutive group of four orbital cycles. Other details are the same as in Figure \ref{fig:gsc_tess37_array2}
\label{fig:gsc_tess64_array2}}
\end{figure*}

\begin{figure*}
\centering
\includegraphics[width=.8\textwidth]{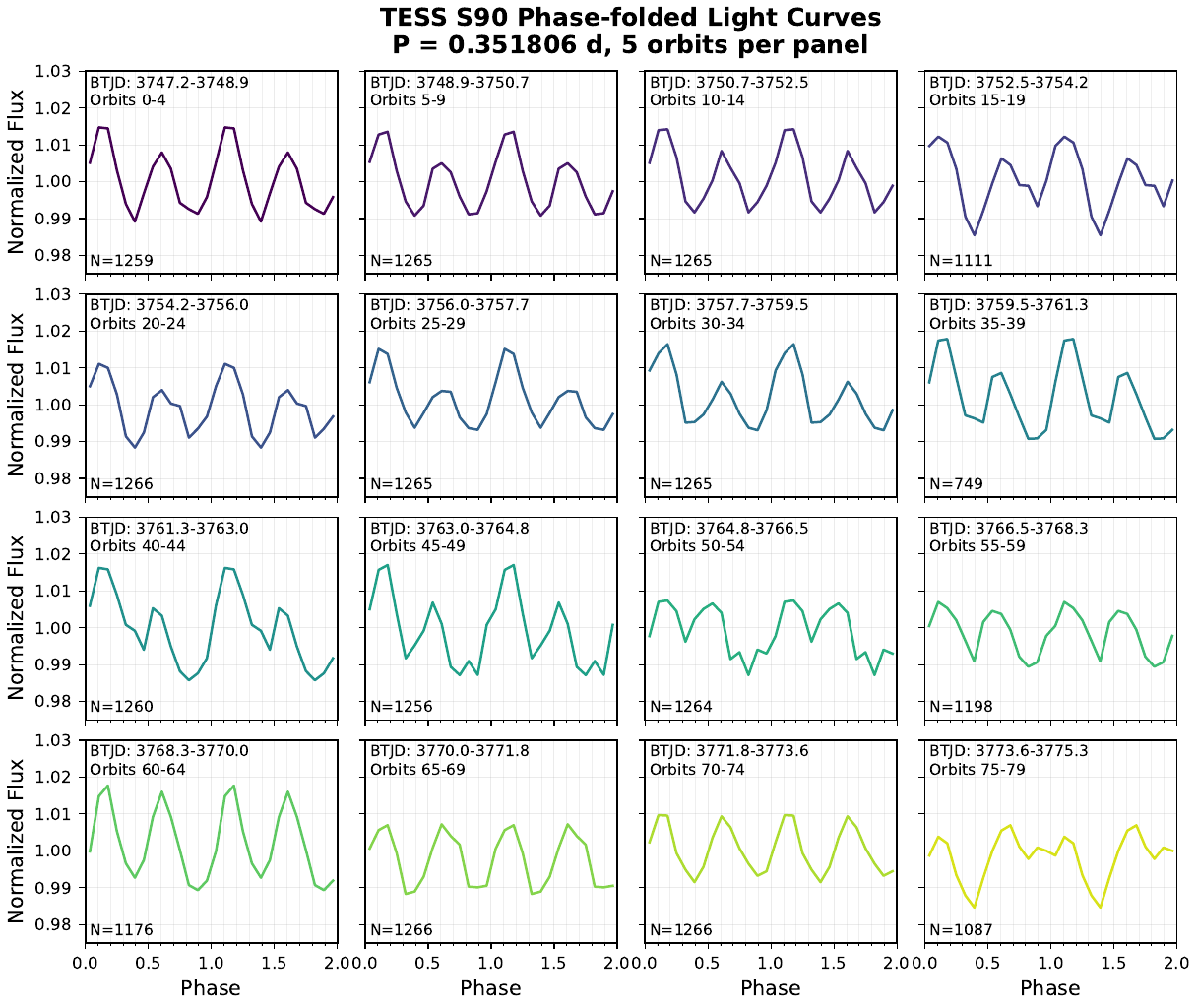}
\caption{Phase-folded light curves of GSC~08227-00723 in TESS Sector 90. Each panel corresponds to a consecutive group of four orbital cycles. Other details are the same as in Figure \ref{fig:gsc_tess37_array2}
\label{fig:gsc_tess90_array2}}
\end{figure*}

\bibliography{RAA-2026-0160}

@article{Kochanek_2017,
   title={The All-Sky Automated Survey for Supernovae (ASAS-SN) Light Curve Server v1.0},
   volume={129},
   ISSN={1538-3873},
   url={http://dx.doi.org/10.1088/1538-3873/aa80d9},
   DOI={10.1088/1538-3873/aa80d9},
   number={980},
   journal={\pasp},
   publisher={IOP Publishing},
   author={Kochanek, C. S. and Shappee, B. J. and Stanek, K. Z. and Holoien, T. W.-S. and Thompson, Todd A. and Prieto, J. L. and Dong, Subo and Shields, J. V. and Will, D. and Britt, C. and et al.},
   year={2017},
   month={Aug},
   pages={104502}
}

@article{10.1117/1.JATIS.1.1.014003,
author = {George R. Ricker and Joshua N. Winn and Roland Vanderspek and David W. Latham and Gáspár  Á. Bakos and Jacob L. Bean and Zachory K. Berta-Thompson and Timothy M. Brown and Lars Buchhave and Nathaniel R. Butler and R. Paul Butler and William J. Chaplin and David B. Charbonneau and Jørgen Christensen-Dalsgaard and Mark Clampin and Drake Deming and John P. Doty and Nathan De Lee and Courtney Dressing and Edward W. Dunham and Michael Endl and François Fressin and Jian Ge and Thomas Henning and Matthew J. Holman and Andrew W. Howard and Shigeru Ida and Jon M. Jenkins and Garrett Jernigan and John Asher Johnson and Lisa Kaltenegger and Nobuyuki Kawai and Hans Kjeldsen and Gregory Laughlin and Alan M. Levine and Douglas Lin and Jack J. Lissauer and Phillip MacQueen and Geoffrey Marcy and Peter R. McCullough and Timothy D. Morton and Norio Narita and Martin Paegert and Enric Palle and Francesco Pepe and Joshua Pepper and Andreas Quirrenbach and Stephen A. Rinehart and Dimitar Sasselov and Bun’ei Sato and Sara Seager and Alessandro Sozzetti and Keivan G. Stassun and Peter Sullivan and Andrew Szentgyorgyi and Guillermo Torres and Stephane Udry and Joel Villasenor},
title = {{Transiting Exoplanet Survey Satellite}},
volume = {1},
journal = {Journal of Astronomical Telescopes, Instruments, and Systems},
number = {1},
publisher = {SPIE},
pages = {1 -- 10},
year = {2014},
doi = {10.1117/1.JATIS.1.1.014003},
URL = {https://doi.org/10.1117/1.JATIS.1.1.014003}
}

@ARTICLE{2023AJ....165..163C,
       author = {{Canbay}, Remziye and {Bilir}, Sel{\c{c}}uk and {{\"O}zd{\"o}nmez}, Aykut and {Ak}, Tansel},
        title = "{Galactic Model Parameters and Spatial Density of Cataclysmic Variables in the Gaia Era: New Constraints on Population Models}",
      journal = {\aj},
         year = 2023,
        month = apr,
       volume = {165},
       number = {4},
          eid = {163},
        pages = {163},
          doi = {10.3847/1538-3881/acbead},
archivePrefix = {arXiv},
       eprint = {2302.11568},
 primaryClass = {astro-ph.SR},
       adsurl = {https://ui.adsabs.harvard.edu/abs/2023AJ....165..163C}
}

@ARTICLE{2022MNRAS.514.4718B,
       author = {{Bruch}, Albert},
        title = "{TESS light curves of cataclysmic variables - I - Unknown periods in long-known stars}",
      journal = {\mnras},
         year = 2022,
        month = aug,
       volume = {514},
       number = {4},
        pages = {4718-4735},
          doi = {10.1093/mnras/stac1650},
archivePrefix = {arXiv},
       eprint = {2207.08203},
 primaryClass = {astro-ph.SR},
       adsurl = {https://ui.adsabs.harvard.edu/abs/2022MNRAS.514.4718B}
}

@ARTICLE{2025ApJS..279...48B,
       author = {{Bruch}, Albert},
        title = "{TESS Light Curves of Cataclysmic Variables. VI. Intermediate Polars}",
      journal = {\apjs},
         year = 2025,
        month = aug,
       volume = {279},
       number = {2},
          eid = {48},
        pages = {48},
          doi = {10.3847/1538-4365/addf41},
       adsurl = {https://ui.adsabs.harvard.edu/abs/2025ApJS..279...48B}
}

@ARTICLE{2023AJ....166...56L,
       author = {{Li}, Xin and {Wang}, Xiaofeng and {Liu}, Jiren and {Guo}, Jincheng and {Zhang}, Ziping and {Sun}, Yongkang and {Song}, Xuan and {Liu}, Cheng},
        title = "{LAMOST J2043+3413-a Fast Disk Precession SW Sextans Candidate in Period Gap}",
      journal = {\aj},
         year = 2023,
        month = aug,
       volume = {166},
       number = {2},
          eid = {56},
        pages = {56},
          doi = {10.3847/1538-3881/acdd70},
archivePrefix = {arXiv},
       eprint = {2306.07529},
 primaryClass = {astro-ph.SR},
       adsurl = {https://ui.adsabs.harvard.edu/abs/2023AJ....166...56L}
}

@ARTICLE{2023MNRAS.519..352B,
       author = {{Bruch}, Albert},
        title = "{TESS light curves of cataclysmic variables - II - Superhumps in old novae and novalike variables}",
      journal = {\mnras},
         year = 2023,
        month = feb,
       volume = {519},
       number = {1},
        pages = {352-376},
          doi = {10.1093/mnras/stac3493},
archivePrefix = {arXiv},
       eprint = {2212.04424},
 primaryClass = {astro-ph.SR},
       adsurl = {https://ui.adsabs.harvard.edu/abs/2023MNRAS.519..352B}
}

@ARTICLE{2023MNRAS.525.1953B,
       author = {{Bruch}, Albert},
        title = "{TESS light curves of cataclysmic variables - III - More superhump systems among old novae and nova-like variables}",
      journal = {\mnras},
         year = 2023,
        month = oct,
       volume = {525},
       number = {2},
        pages = {1953-1975},
          doi = {10.1093/mnras/stad2089},
archivePrefix = {arXiv},
       eprint = {2308.16106},
 primaryClass = {astro-ph.SR},
       adsurl = {https://ui.adsabs.harvard.edu/abs/2023MNRAS.525.1953B}
}

@ARTICLE{2024ApJS..273....6B,
       author = {{Bruch}, Albert},
        title = "{TESS Light Curves of Cataclysmic Variables. IV. A Synoptic View of Eclipsing Old Novae and Novalike Variables}",
      journal = {\apjs},
         year = 2024,
        month = jul,
       volume = {273},
       number = {1},
          eid = {6},
        pages = {6},
          doi = {10.3847/1538-4365/ad43ec},
       adsurl = {https://ui.adsabs.harvard.edu/abs/2024ApJS..273....6B}
}

@ARTICLE{2024AJ....168..121B,
       author = {{Bruch}, Albert},
        title = "{TESS Light Curves of Cataclysmic Variables. V. Improved or Corrected Orbital Periods of 53 Systems}",
      journal = {\aj},
         year = 2024,
        month = sep,
       volume = {168},
       number = {3},
          eid = {121},
        pages = {121},
          doi = {10.3847/1538-3881/ad6260},
       adsurl = {https://ui.adsabs.harvard.edu/abs/2024AJ....168..121B}
}

@book{warner_1995, 
        place={Cambridge}, 
        series={Cambridge Astrophysics}, 
        title={Cataclysmic Variable Stars}, 
        DOI={10.1017/CBO9780511586491}, 
        publisher={Cambridge University Press}, 
        author={Warner, Brian}, 
        year={1995}, 
        collection={Cambridge Astrophysics}}

@article{Shappee_2014,
doi = {10.1088/0004-637X/788/1/48},
url = {https://doi.org/10.1088/0004-637X/788/1/48},
year = {2014},
month = {may},
publisher = {The Astrophysical Journal},
volume = {788},
number = {1},
pages = {48},
author = {Shappee, B. J. and Prieto, J. L. and Grupe, D. and Kochanek, C. S. and Stanek, K. Z. and De Rosa, G. and Mathur, S. and Zu, Y. and Peterson, B. M. and Pogge, R. W. and Komossa, S. and Im, M. and Jencson, J. and Holoien, T.W-S. and Basu, U. and Beacom, J. F. and Szczygieł, D. M. and Brimacombe, J. and Adams, S. and Campillay, A. and Choi, C. and Contreras, C. and Dietrich, M. and Dubberley, M. and Elphick, M. and Foale, S. and Giustini, M. and Gonzalez, C. and Hawkins, E. and Howell, D. A. and Hsiao, E. Y. and Koss, M. and Leighly, K. M. and Morrell, N. and Mudd, D. and Mullins, D. and Nugent, J. M. and Parrent, J. and Phillips, M. M. and Pojmanski, G. and Rosing, W. and Ross, R. and Sand, D. and Terndrup, D. M. and Valenti, S. and Walker, Z. and Yoon, Y.},
title = {THE MAN BEHIND THE CURTAIN: X-RAYS DRIVE THE UV THROUGH NIR VARIABILITY IN THE 2013 ACTIVE GALACTIC NUCLEUS OUTBURST IN NGC 2617},
journal = {\apj}
}

@MISC{2018ascl.soft12013L,
   author = {{Lightkurve Collaboration} and {Cardoso}, J.~V.~d.~M. and
             {Hedges}, C. and {Gully-Santiago}, M. and {Saunders}, N. and
             {Cody}, A.~M. and {Barclay}, T. and {Hall}, O. and
             {Sagear}, S. and {Turtelboom}, E. and {Zhang}, J. and
             {Tzanidakis}, A. and {Mighell}, K. and {Coughlin}, J. and
             {Bell}, K. and {Berta-Thompson}, Z. and {Williams}, P. and
             {Dotson}, J. and {Barentsen}, G.},
    title = "{Lightkurve: Kepler and TESS time series analysis in Python}",
howpublished = {Astrophysics Source Code Library},
     year = 2018,
    month = dec,
archivePrefix = "ascl",
   eprint = {1812.013},
   adsurl = {http://adsabs.harvard.edu/abs/2018ascl.soft12013L},
}

@article{10.1093/mnras/stad2018,
    author = {Inight, Keith and Gänsicke, Boris T and Breedt, Elmé and Israel, Henry T and Littlefair, Stuart P and Manser, Christopher J and Marsh, Tom R and Mulvany, Tim and Pala, Anna Francesca and Thorstensen, John R},
    title = {A catalogue of cataclysmic variables from 20 yr of the Sloan Digital Sky Survey with new classifications, periods, trends, and oddities},
    journal = {Monthly Notices of the Royal Astronomical Society},
    volume = {524},
    number = {4},
    pages = {4867-4898},
    year = {2023},
    month = {07},
    issn = {0035-8711},
    doi = {10.1093/mnras/stad2018},
    url = {https://doi.org/10.1093/mnras/stad2018},
    eprint = {https://academic.oup.com/mnras/article-pdf/524/4/4867/56261577/stad2018.pdf},
}

@ARTICLE{2012RAA....12.1197C,
       author = {{Cui}, Xiang-Qun and {Zhao}, Yong-Heng and {Chu}, Yao-Quan and {Li}, Guo-Ping and {Li}, Qi and {Zhang}, Li-Ping and {Su}, Hong-Jun and {Yao}, Zheng-Qiu and {Wang}, Ya-Nan and {Xing}, Xiao-Zheng and {Li}, Xin-Nan and {Zhu}, Yong-Tian and {Wang}, Gang and {Gu}, Bo-Zhong and {Luo}, A.-Li and {Xu}, Xin-Qi and {Zhang}, Zhen-Chao and {Liu}, Gen-Rong and {Zhang}, Hao-Tong and {Yang}, De-Hua and {Cao}, Shu-Yun and {Chen}, Hai-Yuan and {Chen}, Jian-Jun and {Chen}, Kun-Xin and {Chen}, Ying and {Chu}, Jia-Ru and {Feng}, Lei and {Gong}, Xue-Fei and {Hou}, Yong-Hui and {Hu}, Hong-Zhuan and {Hu}, Ning-Sheng and {Hu}, Zhong-Wen and {Jia}, Lei and {Jiang}, Fang-Hua and {Jiang}, Xiang and {Jiang}, Zi-Bo and {Jin}, Ge and {Li}, Ai-Hua and {Li}, Yan and {Li}, Ye-Ping and {Liu}, Guan-Qun and {Liu}, Zhi-Gang and {Lu}, Wen-Zhi and {Mao}, Yin-Dun and {Men}, Li and {Qi}, Yong-Jun and {Qi}, Zhao-Xiang and {Shi}, Huo-Ming and {Tang}, Zheng-Hong and {Tao}, Qing-Sheng and {Wang}, Da-Qi and {Wang}, Dan and {Wang}, Guo-Min and {Wang}, Hai and {Wang}, Jia-Ning and {Wang}, Jian and {Wang}, Jian-Ling and {Wang}, Jian-Ping and {Wang}, Lei and {Wang}, Shu-Qing and {Wang}, You and {Wang}, Yue-Fei and {Xu}, Ling-Zhe and {Xu}, Yan and {Yang}, Shi-Hai and {Yu}, Yong and {Yuan}, Hui and {Yuan}, Xiang-Yan and {Zhai}, Chao and {Zhang}, Jing and {Zhang}, Yan-Xia and {Zhang}, Yong and {Zhao}, Ming and {Zhou}, Fang and {Zhou}, Guo-Hua and {Zhu}, Jie and {Zou}, Si-Cheng},
        title = "{The Large Sky Area Multi-Object Fiber Spectroscopic Telescope (LAMOST)}",
      journal = {Research in Astronomy and Astrophysics},
         year = 2012,
        month = sep,
       volume = {12},
       number = {9},
        pages = {1197-1242},
          doi = {10.1088/1674-4527/12/9/003},
       adsurl = {https://ui.adsabs.harvard.edu/abs/2012RAA....12.1197C}
}

@ARTICLE{2012RAA....12..723Z,
       author = {{Zhao}, Gang and {Zhao}, Yong-Heng and {Chu}, Yao-Quan and {Jing}, Yi-Peng and {Deng}, Li-Cai},
        title = "{LAMOST spectral survey {\textemdash} An overview}",
      journal = {Research in Astronomy and Astrophysics},
         year = 2012,
        month = jul,
       volume = {12},
       number = {7},
        pages = {723-734},
          doi = {10.1088/1674-4527/12/7/002},
       adsurl = {https://ui.adsabs.harvard.edu/abs/2012RAA....12..723Z}
}

@ARTICLE{2024MNRAS.531..422S,
       author = {{Sun}, Yongkang and {Li}, Xin and {Ao}, Qige and {Cui}, Wenyuan and {Zhang}, Bowen and {Huang}, Yang and {Shi}, Jianrong and {Li}, Linlin and {Liu}, Jifeng},
        title = "{Discovery of a new IW And-type dwarf nova with both tilted disc and tidal instability}",
      journal = {\mnras},
         year = 2024,
        month = jun,
       volume = {531},
       number = {1},
        pages = {422-433},
          doi = {10.1093/mnras/stae1025},
archivePrefix = {arXiv},
       eprint = {2404.09124},
 primaryClass = {astro-ph.SR},
       adsurl = {https://ui.adsabs.harvard.edu/abs/2024MNRAS.531..422S}
}

@ARTICLE{2024ApJ...977..153B,
       author = {{Bruch}, Albert},
        title = "{The AH Pictoris Syndrome: Continuous Trains of Stunted Outbursts in Novalike Variables}",
      journal = {\apj},
         year = 2024,
        month = dec,
       volume = {977},
       number = {2},
          eid = {153},
        pages = {153},
          doi = {10.3847/1538-4357/ad8c39},
archivePrefix = {arXiv},
       eprint = {2410.20756},
 primaryClass = {astro-ph.SR},
       adsurl = {https://ui.adsabs.harvard.edu/abs/2024ApJ...977..153B}
}

@ARTICLE{2001NewAR..45..449L,
       author = {{Lasota}, Jean-Pierre},
        title = "{The disc instability model of dwarf novae and low-mass X-ray binary transients}",
      journal = {\nar},
         year = 2001,
        month = jun,
       volume = {45},
       number = {7},
        pages = {449-508},
          doi = {10.1016/S1387-6473(01)00112-9},
archivePrefix = {arXiv},
       eprint = {astro-ph/0102072},
 primaryClass = {astro-ph},
       adsurl = {https://ui.adsabs.harvard.edu/abs/2001NewAR..45..449L}
}

@ARTICLE{2019PASJ...71...20K,
       author = {{Kato}, Taichi},
        title = "{Three Z Camelopardalis-type dwarf novae exhibiting IW Andromedae-type phenomenon}",
      journal = {\pasj},
         year = 2019,
        month = jan,
       volume = {71},
       number = {1},
          eid = {20},
        pages = {20},
          doi = {10.1093/pasj/psy138},
archivePrefix = {arXiv},
       eprint = {1811.05038},
 primaryClass = {astro-ph.SR},
       adsurl = {https://ui.adsabs.harvard.edu/abs/2019PASJ...71...20K}
}

@article{GaiaDR3,
  author = {{Gaia Collaboration} and Vallenari, A. and Brown, A.~G.~A. and others},
  title = {Gaia Data Release 3: Summary of the content and survey properties},
  journal = {Astronomy \& Astrophysics},
  volume = {674},
  pages = {A1},
  year = {2023},
  doi = {10.1051/0004-6361/202243940}
}

@article{Savitzky1964,
  author = {Savitzky, Abraham and Golay, Marcel J. E.},
  title = {Smoothing and Differentiation of Data by Simplified Least Squares Procedures},
  journal = {Analytical Chemistry},
  volume = {36},
  pages = {1627--1639},
  year = {1964},
  doi = {10.1021/ac60214a047}
}

@article{Lomb1976,
  author  = {Lomb, N. R.},
  title   = {Least-Squares Frequency Analysis of Unequally Spaced Data},
  journal = {Astrophysics and Space Science},
  volume  = {39},
  pages   = {447--462},
  year    = {1976},
  doi     = {10.1007/BF00648343}
}

@article{Scargle1982,
  author  = {Scargle, J. D.},
  title   = {Studies in Astronomical Time Series Analysis. II. Statistical Aspects of Spectral Analysis of Unevenly Spaced Data},
  journal = {\apj},
  volume  = {263},
  pages   = {835--853},
  year    = {1982},
  doi     = {10.1086/160554}
}

@article{Breger1993,
  author  = {Breger, M. and Stich, J. and Garrido, R. and others},
  title   = {Nonradial Pulsation of the Delta Scuti Star FG Virginis},
  journal = {\aap},
  volume  = {271},
  pages   = {482--501},
  year    = {1993}
}

@article{VanderPlas2018,
  author  = {VanderPlas, J. T.},
  title   = {Understanding the Lomb--Scargle Periodogram},
  journal = {The Astrophysical Journal Supplement Series},
  volume  = {236},
  pages   = {16},
  year    = {2018},
  doi     = {10.3847/1538-4365/aab766}
}

@misc{DSS2,
  author = {STScI},
  title = {Digitized Sky Survey 2},
  howpublished = {\url{http://archive.eso.org/dss/dss/}},
  year = {1996}
}

@ARTICLE{2001MNRAS.325...89C,
       author = {{Chen}, A. and {O'Donoghue}, D. and {Stobie}, R.~S. and {Kilkenny}, D. and {Warner}, B.},
        title = "{Cataclysmic variables in the Edinburgh-Cape Blue Object SurveyQ3}",
      journal = {\mnras},
         year = 2001,
        month = jul,
       volume = {325},
       number = {1},
        pages = {89-110},
          doi = {10.1046/j.1365-8711.2001.04322.x},
       adsurl = {https://ui.adsabs.harvard.edu/abs/2001MNRAS.325...89C}
}

@ARTICLE{2025ApJS..277...29H,
       author = {{Honeycutt}, R.~K. and {Robertson}, Jeff W. and {Radzom}, Brandon T.},
        title = "{Stunted Outbursts and Z Cam{\textendash}like Behaviors in the Long-term Light Curves of Novalike Cataclysmic Variables}",
      journal = {\apjs},
         year = 2025,
        month = mar,
       volume = {277},
       number = {1},
          eid = {29},
        pages = {29},
          doi = {10.3847/1538-4365/adad75},
       adsurl = {https://ui.adsabs.harvard.edu/abs/2025ApJS..277...29H}
}

@ARTICLE{2009MNRAS.398.2110W,
       author = {{Wood}, M.~A. and {Thomas}, D.~M. and {Simpson}, J.~C.},
        title = "{SPH simulations of negative (nodal) superhumps: a parametric study}",
      journal = {\mnras},
         year = 2009,
        month = oct,
       volume = {398},
       number = {4},
        pages = {2110-2121},
          doi = {10.1111/j.1365-2966.2009.15252.x},
archivePrefix = {arXiv},
       eprint = {0906.2713},
 primaryClass = {astro-ph.SR},
       adsurl = {https://ui.adsabs.harvard.edu/abs/2009MNRAS.398.2110W}
}

@ARTICLE{2019ApJ...887...93G,
       author = {{Green}, Gregory M. and {Schlafly}, Edward and {Zucker}, Catherine and {Speagle}, Joshua S. and {Finkbeiner}, Douglas},
        title = "{A 3D Dust Map Based on Gaia, Pan-STARRS 1, and 2MASS}",
      journal = {\apj},
         year = 2019,
        month = dec,
       volume = {887},
       number = {1},
          eid = {93},
        pages = {93},
          doi = {10.3847/1538-4357/ab5362},
archivePrefix = {arXiv},
       eprint = {1905.02734},
 primaryClass = {astro-ph.GA},
       adsurl = {https://ui.adsabs.harvard.edu/abs/2019ApJ...887...93G}
}

@article{Szkody_2013,
doi = {10.1086/674170},
url = {https://doi.org/10.1086/674170},
year = {2013},
month = {dec},
publisher = {University of Chicago Press},
volume = {125},
number = {934},
pages = {1421},
author = {Szkody, Paula and Albright, Meagan and Linnell, Albert P. and Everett, Mark E. and McMillan, Russet and Saurage, Gabrelle and Huehnerhoff, Joseph and Howell, Steve B. and Simonsen, Mike and Hunt-Walker, Nick},
title = {A Study of the Unusual Z Cam Systems IW Andromedae and V513 Cassiopeia1},
journal = {Publications of the Astronomical Society of the Pacific}
}

@article{10.1093/pasj/psaa096,
    author = {Kato, Taichi and Kojiguchi, Naoto},
    title = {BC Cassiopeiae: First detection of IW Andromedae-type phenomenon among post-eruption novae},
    journal = {\pasj},
    volume = {72},
    number = {6},
    pages = {98},
    year = {2020},
    month = {10},
    issn = {0004-6264},
    doi = {10.1093/pasj/psaa096},
    url = {https://doi.org/10.1093/pasj/psaa096},
    eprint = {https://academic.oup.com/pasj/article-pdf/72/6/98/54663803/psaa096.pdf},
}

@article{10.1093/pasj/psab074,
    author = {Kato, Taichi and Tampo, Yusuke and Kojiguchi, Naoto and Shibata, Masaaki and Ito, Junpei and Isogai, Keisuke and Itoh, Hiroshi and Hambsch, Franz-Josef and Monard, Berto and Kiyota, Seiichiro and Vanmunster, Tonny and Sosnovskij, Aleksei A and Pavlenko, Elena P and Dubovsky, Pavol A and Kudzej, Igor and Medulka, Tomas},
    title = {BO Ceti: Dwarf nova showing both IW And-type and SU UMa-Type features},
    journal = {\pasj},
    volume = {73},
    number = {5},
    pages = {1280-1288},
    year = {2021},
    month = {07},
    issn = {0004-6264},
    doi = {10.1093/pasj/psab074},
    url = {https://doi.org/10.1093/pasj/psab074},
    eprint = {https://academic.oup.com/pasj/article-pdf/73/5/1280/54654820/psab074.pdf},
}

@ARTICLE{2014JAVSO..42..199S,
       author = {{Simonsen}, M. and {Bohlsen}, T. and {Hambsch}, F.-J. and {Stubbings}, R.},
        title = "{ST Chamaeleontis and BP Coronae Australis: Two Southern Dwarf Novae Confirmed as Z Cam Stars}",
      journal = {JAAVSO},
         year = 2014,
        month = jun,
       volume = {42},
       number = {1},
        pages = {199},
          doi = {10.48550/arXiv.1402.0210},
archivePrefix = {arXiv},
       eprint = {1402.0210},
 primaryClass = {astro-ph.SR},
       adsurl = {https://ui.adsabs.harvard.edu/abs/2014JAVSO..42..199S}
}

@article{10.1093/pasj/psy138,
    author = {Kato, Taichi},
    title = {Three Z Camelopardalis-type dwarf novae exhibiting IW Andromedae-type phenomenon},
    journal = {\pasj},
    volume = {71},
    number = {1},
    pages = {20},
    year = {2018},
    month = {12},
    issn = {0004-6264},
    doi = {10.1093/pasj/psy138},
    url = {https://doi.org/10.1093/pasj/psy138},
    eprint = {https://academic.oup.com/pasj/article-pdf/71/1/20/54665545/pasj_71_1_20.pdf},
}

@ARTICLE{2022arXiv220305143K,
       author = {{Kato}, Taichi and {Moriyama}, Masayuki},
        title = "{Long-lasting standstill and fading episode in the IW And star V507 Cyg}",
      journal = {arXiv e-prints},
         year = 2022,
        month = mar,
          eid = {arXiv:2203.05143},
        pages = {arXiv:2203.05143},
          doi = {10.48550/arXiv.2203.05143},
archivePrefix = {arXiv},
       eprint = {2203.05143},
 primaryClass = {astro-ph.SR},
       adsurl = {https://ui.adsabs.harvard.edu/abs/2022arXiv220305143K}
}

@ARTICLE{2016A&A...589A.106M,
       author = {{Mason}, E. and {Howell}, S.~B.},
        title = "{Kepler and Hale observations of V523 Lyrae}",
      journal = {\aap},
         year = 2016,
        month = may,
       volume = {589},
          eid = {A106},
        pages = {A106},
          doi = {10.1051/0004-6361/201628245},
archivePrefix = {arXiv},
       eprint = {1603.01410},
 primaryClass = {astro-ph.SR},
       adsurl = {https://ui.adsabs.harvard.edu/abs/2016A&A...589A.106M}
}

@ARTICLE{2021ApJ...911...51L,
       author = {{Lee}, Chien-De and {Ou}, Jia-Yu and {Yu}, Po-Chieh and {Ngeow}, Chow-Choong and {Huang}, Po-Chieh and {Ip}, Wing-Huen and {Hambsch}, Franz-Josef and {Sung}, Hyun-il and {van Roestel}, Jan and {Dekany}, Richard and {Drake}, Andrew J. and {Graham}, Matthew J. and {Duev}, Dmitry A. and {Kaye}, Stephen and {Kupfer}, Thomas and {Laher}, Russ R. and {Masci}, Frank J. and {Mr{\'o}z}, Przemek and {Neill}, James D. and {Riddle}, Reed and {Rusholme}, Ben and {Walters}, Richard},
        title = "{HO Puppis: Not a Be Star, but a Newly Confirmed IW And-type Star}",
      journal = {\apj},
         year = 2021,
        month = apr,
       volume = {911},
       number = {1},
          eid = {51},
        pages = {51},
          doi = {10.3847/1538-4357/abe871},
archivePrefix = {arXiv},
       eprint = {2102.09748},
 primaryClass = {astro-ph.SR},
       adsurl = {https://ui.adsabs.harvard.edu/abs/2021ApJ...911...51L}
}

@ARTICLE{2013ApJ...775...64G,
       author = {{Gies}, Douglas R. and {Guo}, Zhao and {Howell}, Steve B. and {Still}, Martin D. and {Boyajian}, Tabetha S. and {Hoekstra}, Abe J. and {Jek}, Kian J. and {LaCourse}, Daryll and {Winarski}, Troy},
        title = "{KIC 9406652: An Unusual Cataclysmic Variable in the Kepler Field of View}",
      journal = {\apj},
         year = 2013,
        month = sep,
       volume = {775},
       number = {1},
          eid = {64},
        pages = {64},
          doi = {10.1088/0004-637X/775/1/64},
archivePrefix = {arXiv},
       eprint = {1308.0369},
 primaryClass = {astro-ph.SR},
       adsurl = {https://ui.adsabs.harvard.edu/abs/2013ApJ...775...64G}
}

@ARTICLE{2020PASJ...72...94K,
       author = {{Kimura}, Mariko and {Osaki}, Yoji and {Kato}, Taichi},
        title = "{KIC 9406652: A laboratory for tilted disks in cataclysmic variable stars}",
      journal = {\pasj},
         year = 2020,
        month = dec,
       volume = {72},
       number = {6},
          eid = {94},
        pages = {94},
          doi = {10.1093/pasj/psaa088},
archivePrefix = {arXiv},
       eprint = {2008.11328},
 primaryClass = {astro-ph.SR},
       adsurl = {https://ui.adsabs.harvard.edu/abs/2020PASJ...72...94K}
}

@ARTICLE{2024ApJ...976..107S,
       author = {{Sun}, Qi-Bin and {Qian}, Sheng-Bang and {Zhu}, Li-Ying and {Li}, Qin-Mei and {Li}, Fu-Xing and {Li}, Min-Yu and {Li}, Ping},
        title = "{A New IW And-type Star: Karachurin 12 with Tilted Disks and Diverse Cycles}",
      journal = {\apj},
         year = 2024,
        month = nov,
       volume = {976},
       number = {1},
          eid = {107},
        pages = {107},
          doi = {10.3847/1538-4357/ad8446},
archivePrefix = {arXiv},
       eprint = {2409.03011},
 primaryClass = {astro-ph.SR},
       adsurl = {https://ui.adsabs.harvard.edu/abs/2024ApJ...976..107S}
}

@ARTICLE{1991PASP..103..735P,
       author = {{Patterson}, Joseph and {Richman}, Hayley},
        title = "{Permanent Superhumps in V603 Aquilae}",
      journal = {\pasp},
         year = 1991,
        month = aug,
       volume = {103},
        pages = {735},
          doi = {10.1086/132872},
       adsurl = {https://ui.adsabs.harvard.edu/abs/1991PASP..103..735P}
}

@ARTICLE{1998PASP..110.1132P,
       author = {{Patterson}, Joseph},
        title = "{Late Evolution of Cataclysmic Variables}",
      journal = {\pasp},
         year = 1998,
        month = oct,
       volume = {110},
       number = {752},
        pages = {1132-1147},
          doi = {10.1086/316233},
       adsurl = {https://ui.adsabs.harvard.edu/abs/1998PASP..110.1132P}
}

@ARTICLE{1988MNRAS.232...35W,
       author = {{Whitehurst}, Robert},
        title = "{Numerical simulations of accretion discs - I. Superhumps : a tidal phenomenon of accretion discs.}",
      journal = {\mnras},
         year = 1988,
        month = may,
       volume = {232},
        pages = {35-51},
          doi = {10.1093/mnras/232.1.35},
       adsurl = {https://ui.adsabs.harvard.edu/abs/1988MNRAS.232...35W}
}

@ARTICLE{1996PASP..108...39O,
       author = {{Osaki}, Yoji},
        title = "{Dwarf-Nova Outbursts}",
      journal = {\pasp},
         year = 1996,
        month = jan,
       volume = {108},
        pages = {39},
          doi = {10.1086/133689},
       adsurl = {https://ui.adsabs.harvard.edu/abs/1996PASP..108...39O}
}

@ARTICLE{2005PASP..117.1204P,
       author = {{Patterson}, Joseph and {Kemp}, Jonathan and {Harvey}, David A. and {Fried}, Robert E. and {Rea}, Robert and {Monard}, Berto and {Cook}, Lewis M. and {Skillman}, David R. and {Vanmunster}, Tonny and {Bolt}, Greg and {Armstrong}, Eve and {McCormick}, Jennie and {Krajci}, Thomas and {Jensen}, Lasse and {Gunn}, Jerry and {Butterworth}, Neil and {Foote}, Jerry and {Bos}, Marc and {Masi}, Gianluca and {Warhurst}, Paul},
        title = "{Superhumps in Cataclysmic Binaries. XXV. q$_{crit}$, ɛ(q), and Mass-Radius}",
      journal = {\pasp},
         year = 2005,
        month = nov,
       volume = {117},
       number = {837},
        pages = {1204-1222},
          doi = {10.1086/447771},
archivePrefix = {arXiv},
       eprint = {astro-ph/0507371},
 primaryClass = {astro-ph},
       adsurl = {https://ui.adsabs.harvard.edu/abs/2005PASP..117.1204P}
}

@ARTICLE{1989PASJ...41.1005O,
       author = {{Osaki}, Yoji},
        title = "{A Model for the Superoutburst Phenomenon of SU Ursae Majoris Stars}",
      journal = {\pasj},
         year = 1989,
        month = dec,
       volume = {41},
       number = {5},
        pages = {1005-1033},
          doi = {10.1093/pasj/41.5.1005},
       adsurl = {https://ui.adsabs.harvard.edu/abs/1989PASJ...41.1005O}
}

@ARTICLE{2003MNRAS.340..679R,
       author = {{Retter}, A. and {Hellier}, C. and {Augusteijn}, T. and {Naylor}, T. and {Bedding}, T.~R. and {Bembrick}, C. and {McCormick}, J. and {Velthuis}, F.},
        title = "{A 6.3-h superhump in the cataclysmic variable TV Columbae: the longest yet seen}",
      journal = {\mnras},
         year = 2003,
        month = apr,
       volume = {340},
       number = {2},
        pages = {679-686},
          doi = {10.1046/j.1365-8711.2003.06331.x},
archivePrefix = {arXiv},
       eprint = {astro-ph/0212151},
 primaryClass = {astro-ph},
       adsurl = {https://ui.adsabs.harvard.edu/abs/2003MNRAS.340..679R}
}

@ARTICLE{2022Natur.604..447S,
       author = {{Scaringi}, S. and {Groot}, P.~J. and {Knigge}, C. and {Bird}, A.~J. and {Breedt}, E. and {Buckley}, D.~A.~H. and {Cavecchi}, Y. and {Degenaar}, N.~D. and {de Martino}, D. and {Done}, C. and {Fratta}, M. and {I{\l}kiewicz}, K. and {Koerding}, E. and {Lasota}, J.-P. and {Littlefield}, C. and {Manara}, C.~F. and {O'Brien}, M. and {Szkody}, P. and {Timmes}, F.~X.},
        title = "{Localized thermonuclear bursts from accreting magnetic white dwarfs}",
      journal = {\nat},
         year = 2022,
        month = apr,
       volume = {604},
       number = {7906},
        pages = {447-450},
          doi = {10.1038/s41586-022-04495-6},
archivePrefix = {arXiv},
       eprint = {2204.09070},
 primaryClass = {astro-ph.HE},
       adsurl = {https://ui.adsabs.harvard.edu/abs/2022Natur.604..447S}
}

@ARTICLE{1998ApJ...500..525S,
       author = {{Schlegel}, David J. and {Finkbeiner}, Douglas P. and {Davis}, Marc},
        title = "{Maps of Dust Infrared Emission for Use in Estimation of Reddening and Cosmic Microwave Background Radiation Foregrounds}",
      journal = {\apj},
         year = 1998,
        month = jun,
       volume = {500},
       number = {2},
        pages = {525-553},
          doi = {10.1086/305772},
archivePrefix = {arXiv},
       eprint = {astro-ph/9710327},
 primaryClass = {astro-ph},
       adsurl = {https://ui.adsabs.harvard.edu/abs/1998ApJ...500..525S}
}

@ARTICLE{1991ApJ...381..259L,
       author = {{Lubow}, Stephen H.},
        title = "{A Model for Tidally Driven Eccentric Instabilities in Fluid Disks}",
      journal = {\apj},
         year = 1991,
        month = nov,
       volume = {381},
        pages = {259},
          doi = {10.1086/170647},
       adsurl = {https://ui.adsabs.harvard.edu/abs/1991ApJ...381..259L}
}

@INPROCEEDINGS{2008AIPC.1054..101W,
       author = {{Warner}, Brian and {Woudt}, Patrick A.},
        title = "{QPOs in CVs: An executive summary}",
    booktitle = {Cool Discs, Hot Flows: The Varying Faces of Accreting Compact Objects},
         year = 2008,
       editor = {{Axelsson}, Magnus},
       series = {American Institute of Physics Conference Series},
       volume = {1054},
        month = sep,
    publisher = {AIP},
        pages = {101-110},
          doi = {10.1063/1.3002491},
archivePrefix = {arXiv},
       eprint = {0806.1317},
 primaryClass = {astro-ph},
       adsurl = {https://ui.adsabs.harvard.edu/abs/2008AIPC.1054..101W}
}

@ARTICLE{2022MNRAS.510.3605I,
       author = {{Inight}, K. and {G{\"a}nsicke}, B.~T. and {Blondel}, D. and {Boyd}, D. and {Ashley}, R.~P. and {Knigge}, C. and {Long}, K.~S. and {Marsh}, T.~R. and {McCleery}, J. and {Scaringi}, S. and {Steeghs}, D. and {Thorstensen}, J.~R. and {Vanmunster}, T. and {Wheatley}, P.~J.},
        title = "{ASAS J071404+7004.3 - a close, bright nova-like cataclysmic variable with gusty winds}",
      journal = {\mnras},
         year = 2022,
        month = mar,
       volume = {510},
       number = {3},
        pages = {3605-3621},
          doi = {10.1093/mnras/stab3662},
archivePrefix = {arXiv},
       eprint = {2109.14514},
 primaryClass = {astro-ph.SR},
       adsurl = {https://ui.adsabs.harvard.edu/abs/2022MNRAS.510.3605I}
}

@article{Warner_2004,
doi = {10.1086/381742},
url = {https://doi.org/10.1086/381742},
year = {2004},
month = {jan},
publisher = {The University of Chicago Press},
volume = {116},
number = {816},
pages = {115},
author = {Warner, Brian},
title = {Rapid Oscillations in Cataclysmic Variables},
journal = {Publications of the Astronomical Society of the Pacific}
}

@article{2011ApJS..194...28K,
  author  = {Knigge, C. and Baraffe, I. and Patterson, J.},
  title   = {The Evolution of Cataclysmic Variables as Revealed by Their Donor Stars},
  journal = {The Astrophysical Journal Supplement Series},
  year    = {2011},
  volume  = {194},
  number  = {2},
  pages   = {28},
  doi     = {10.1088/0067-0049/194/2/28}
}
\bibliographystyle{raa.bst}

\end{document}